\documentclass[aps,pra,10pt, twocolumn,preprintnumbers, superscriptaddress,nofootinbib,longbibliography]{revtex4-2}

\usepackage[colorlinks=true,
            linkcolor=EqRefBlue,   
            citecolor=CiteRed,   
            urlcolor=CiteGreen     
           ]{hyperref}

\usepackage{amsmath}     
\usepackage{mathtools}    
\usepackage{amssymb}
\usepackage{amsthm}
\newtheorem*{theorem*}{Theorem}
\newtheorem{theorem}{Theorem}[section]

\newtheorem{assumption}{Assumption}

\theoremstyle{definition}

\usepackage[normalem]{ulem}

\usepackage{xcolor}

\usepackage[caption=false]{subfig}
\usepackage{graphicx,tikz} 
\usepackage{simpler-wick}
\usepackage{cleveref}
\crefname{assumption}{assumption}{assumptions}
\crefname{definition}{definition}{definitions}
\Crefname{assumption}{Assumption}{Assumptions}
\Crefname{definition}{Definition}{Definitions}

\usetikzlibrary{calc}
\usepackage{braket}
\newcommand{\be}{\begin{equation}}
\newcommand{\ee}{\end{equation}}

\newcommand{\sH}{\mathcal{H}}

\usepackage{xurl}
\usepackage[normalem]{ulem}

\definecolor{CiteRed}{HTML}{C00000} 
\definecolor{EqRefBlue}{HTML}{1F4B99}
\definecolor{CiteGreen}{HTML}{0B7A0B}
\definecolor{EditPurple}{RGB}{128,0,128}

\newcommand{\sS}{\mathcal{S}} 
\newcommand{\inst}{\langle\Psi_{\mathrm{in}}(B_0)|}
\newcommand{\out}{|\Psi_{\mathrm{out}}(M_x)\rangle}

\definecolor{softmaroon}{RGB}{150,60,75}

\usepackage{tabularx}   
\usepackage{booktabs}   
\usepackage{array}      
\newcolumntype{C}{>{\centering\arraybackslash}X} 
\newcolumntype{P}[1]{>{\centering\arraybackslash}p{#1}} 
\newcolumntype{M}[1]{>{\centering\arraybackslash}m{#1}}

\newcommand{\otoc}{{\rm OTOC}}
\newcommand{\id}{{\mathbb{I}}}

\newcommand{\hd}{{\hat{d}}}
\newcommand{\hb}{{\hat{b}}}

\begin{document}

\title{An efficient classical algorithm for OTOCs in chaotic quantum dynamics}

\author{Yunchao Liu}
\affiliation{IBM Research, Cambridge, MA 02142, USA}
\author{Yihui Quek}
\affiliation{School of Computer and Communication Sciences and School of Basic Sciences, École Polytechnique Fédérale de Lausanne}
\affiliation{School of Computing, National University of Singapore}
\author{Shreya Vardhan}
\affiliation{Institute for Quantum Information and Matter, California Institute of Technology, Pasadena, CA 91125, USA}
\author{Tianci Zhou}
\affiliation{Department of Physics, Virginia Tech, Blacksburg, VA 24061, USA}

\begin{abstract}
Estimating out-of-time-ordered correlators (OTOCs) in individual instances of random quantum circuits has been proposed as a candidate for verifiable quantum advantage, and demonstrated to be hard for conventional classical algorithms such as tensor networks and Monte Carlo. Using insights from quantum chaos, we develop and implement a polynomial-time classical algorithm for the four-point $\otoc^{(1)}$ and the eight-point $\otoc^{(2)}$ in typical instances of one-dimensional (1D) quantum circuits. 
Our results demonstrate average-case classical tractability of estimating OTOCs in the sensitive regime where circuit-to-circuit fluctuations are large.  
Our algorithm organizes the OTOC into a rapidly converging series, where the leading term includes contributions from histories within permutation degrees of freedom in the replicated Hilbert space, and subleading terms arise from excursions outside this space. 
Although chaos is often viewed as an insurmountable barrier to classical simulation, our algorithm exploits a form of dephasing caused by chaos, which suppresses the subleading terms. We provide strong theoretical and numerical evidence to support this suppression. Further, we demonstrate  that practical implementations of our algorithm estimate $\otoc^{(1)}$ and $\otoc^{(2)}$ in 1D circuits with high accuracy, achieving higher signal-to-noise ratio at larger circuit depths compared to recent 2D experiments by Google. Extending our algorithm to higher dimensions, and resolving the status of quantum advantage in 2D OTOCs, remain open problems.
\end{abstract}

\maketitle 
\section{Introduction}
Demonstrating quantum advantage is a major milestone on the path towards practical quantum computing. At present, this effort focuses on identifying computational problems that are hard for classical computers, yet solvable on today's noisy intermediate-scale quantum (NISQ) devices. Ideally, these problems should also yield verifiable results and practical applications. Expectation values involving time-evolved observables in quantum many-body systems provide a rich source of such problems. Such quantities are natural candidates for classical hardness, due to the exponentially large quantum many-body Hilbert space explored by the dynamics, as well as the nontrivial effects of quantum interference. Further,  expectation values represent a fundamentally new milestone relative to previously demonstrated sampling tasks~\cite{Arute2019,Zhong2020}, as they can in principle be verified on a trusted quantum computer given the parameters of the initial state and time evolution~\cite{Kim2023,leviatan2026resolvingstructureprethermalfloquet,barron2026observableestimationabsenceclassical}. Computing these observables also allows us to probe physically interesting emergent behaviors of quantum many-body systems on large space and time scales. 

A recent work by Google~\cite{Abanin2025} measured observables known as out-of-time-ordered correlators (OTOCs) in random quantum circuits, and proposed that the computation of these observables provides a candidate for quantum advantage. The $2k$-point OTOC for local  operators $M$ and $B$ is defined as ${\otoc}^{(k)}(t)=\langle [B(t)M]^{2k}\rangle$, where $B(t) = U(t)^{\dagger} B U(t)$ is the time-evolved operator under a unitary evolution $U(t)$, and $\langle\cdot\rangle$ denotes an expectation value in a quantum state, typically taken to be a product state or thermal state. The $k=1$ case yields the standard four-point OTOC, which has been widely studied as a measure of chaos and scrambling in quantum many-body systems and quantum gravity~\cite{otoc_rev_1,otoc_rev_2, otoc_rev_3, otoc_rev_4, nahum_op, vk_op,maldacenaBoundChaos2016,hosurChaosQuantumChannels2016,guTwoWayApproachOutofTimeOrder2022,harrowSeparationOutOfTimeOrderedCorrelation2021,gonzalezalonsoOutOfTimeOrderedCorrelatorQuasiprobabilitiesRobustly2019} (see~\cite{otoc_rev_5} for a review; also see~\cite{cotler, schuster} for applications to Hamiltonian learning). Higher-point OTOCs, corresponding to $k>1$, have recently also been shown to capture more fine-grained features of quantum chaos~\cite{fmi, higher_1,higher_2,higher_3,higher_4,higher_5,higher_6,higher_7,higher_8,higher_9,higher_10,higher_11,higher_12,higher_13,robertsChaosComplexityDesign2017}.  
While the average $\mathbb{E}_U[{\otoc}^{(1)}]$ over random unitary circuits $U$ is well-known to be easily computable~\cite{nahum_op,vk_op,bertiniScramblingRandomUnitary2020},  ${\otoc}^{(k)}$ for an individual circuit realization is believed to be classically hard to compute, especially for $k\geq 2$. To assess this conjectured hardness, several classes of classical algorithms have been considered for this task, including tensor networks, Monte Carlo approaches,  Pauli propagation,  and Clifford expansions \cite{Abanin2025,bermejo2026tensornetworksbeliefpropagation,mele2026verifiablequantumadvantagerandom}. None of these algorithms have yielded efficient simulations of higher-order OTOCs at depths attainable by the experiment of~\cite{Abanin2025}.

In this paper, we present an efficient classical algorithm for ${\otoc}^{(1)}$ and ${\otoc}^{(2)}$ for random quantum circuits in one spatial dimension (1D), and thus demonstrate the classical tractability of this simulation task. More explicitly, our result applies to typical instances of random circuits, in the sensitive regime where OTOC fluctuations between circuits are large. It thus shows 
{\it average-case} classical simulability of OTOCs in this regime, despite the $\textsf{DQC1}$-completeness of this problem in the worst case~\cite{roymoulikDQC1hardnessEstimatingCorrelation2024}. Our algorithm approximates these quantities in an individual circuit through a perturbative expansion around (a subset of)  states appearing in the average over random circuits. We provide theoretical arguments for the rapid convergence of this expansion. By appropriately truncating the expansion, one can achieve any desired inverse polynomial error with polynomial computational resources. We show that our algorithm performs well not only in theory but also in practice,  by demonstrating numerically that it estimates the OTOCs to high accuracy in 1D, at comparable circuit depths to the 2D experiment of~\cite{Abanin2025}.

Many existing efficient classical algorithms for quantum many-body physics, such as tensor network algorithms, often target a polynomial parameterization of the wavefunction. Thus chaos, which is associated with scrambling and entanglement growth in the Hilbert space~\cite{nahumQuantumEntanglementGrowth2017,aleinerMicroscopicModelQuantum2016,xuAccessingScramblingMatrix2020,chenOperatorScramblingQuantum2018,maldacenaBoundChaos2016,hosurChaosQuantumChannels2016,harrowSeparationOutOfTimeOrderedCorrelation2021,xuLocalityQuantumFluctuations2019}, is often seen as an insurmountable barrier to efficient simulation.  As emphasized in~\cite{Abanin2025}, by viewing $U(t)^{\dagger} B U(t)$ in the Pauli basis, the higher-point OTOCs appear to be particularly intractable: they are sensitive both to the coefficients of exponentially many operators, which preclude Pauli-path algorithms~\cite{Rakovszky2022DAOE,Gao2018NoisySimulation,Aharonov2023NoisySampling,Schuster2025NoisyCircuits,GonzalezGarcia2025PauliPaths}, and to fluctuating phases, which preclude classical Monte Carlo methods. However, our algorithm defies this apparent complexity by using an emergent property of chaos, due to which the dynamics simplify in a certain ``permutation basis'', which we will explain below. On viewing the dynamics in this basis, most states in the exponentially large Hilbert space turn out to be irrelevant for the OTOCs.

\begin{figure*}[!t]
\centering
\includegraphics[width=\textwidth]{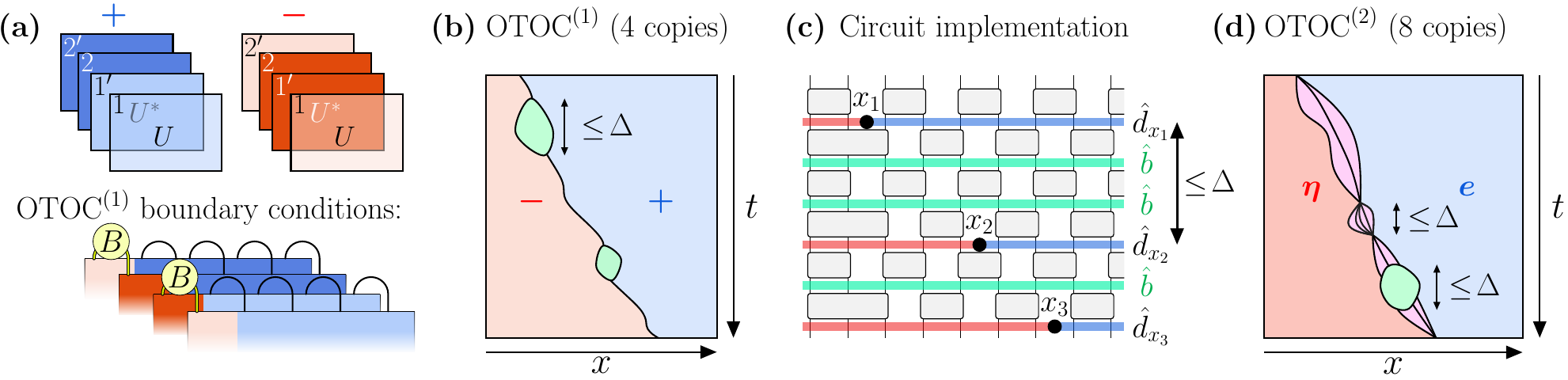}
\caption{{\bf (a)}:
The $+$ and $-$ pairings for two forward and two backward copies of the time-evolution. In $\otoc^{(1)}$, the $B$ operator induces a boundary condition with a domain wall between $\ket{-}$ and $\ket{+}$. Copies with the same color are paired. {\bf (b)}: On viewing $\otoc^{(1)}$ as a transition amplitude, we get a sum over histories where the dominant contributions have a single domain wall between $\ket{-}$ and $\ket{+}$ at most times, with only short excursions (green bubbles) outside this single domain wall subspace.  In the membrane cluster expansion algorithm, we truncate to configurations with bubbles of some maximum duration $\Delta$, which can be chosen depending on the desired accuracy. \textbf{(c):} We show explicitly how the domain wall and bubble configurations are evaluated in the algorithm  on the brickwork circuit geometry, with horizontal lines below the gate layers showing either  projectors onto domain walls, $\hat d$ (red/blue), or projectors  onto the complementary  ``bubble'' subspace (green). $\hat d = \sum_x \hat d_x$, where $\hat d_x$ projects onto a domain wall at position $x$. Hence, both $\hat d$ and the complementary projector are non-local, but we have shown examples of  local terms in the former.  {\bf (d):} $\otoc^{(2)}$ can be expressed as a transition amplitude on 8 copies of the system, under $(U \otimes U^{\ast})^{\otimes 4}$.  The top boundary condition now leads to ``composite'' domain walls containing up to 3 ``elementary'' domain walls. We can have two types of bubble contributions here: one consisting of additional domain walls within the permutation subspace (which give non-trivial contributions even to the random average for this quantity), and another consisting of departures from the permutation subspace. }
\label{intuition_figure}
\end{figure*}

The subspace that remains relevant for the dynamics of the OTOCs is spanned  by {\it few-wall} states with a small number of domain walls or ``membranes'' between distinct permutations~\cite{nahumQuantumEntanglementGrowth2017,zhouEmergentStatisticalMechanics2019,zhou_entanglement_2020}. We build a systematic expansion for corrections on top of these membranes, which we refer to as the ``membrane cluster expansion'' (see Fig.~\ref{intuition_figure} for an intuitive illustration).  Our key physical assumption is the strong suppression of such corrections as a function of their time duration. We expect this assumption to apply in both one and higher spatial dimensions. In 1D, we show strong support for this assumption with a combination of theoretical arguments and numerical tests. We have the further simplification in 1D that the number of few-wall states at a given time slice is polynomial in time, enabling efficient resummation of our perturbative expansion. In higher dimensions, even though we expect that the few-wall states continue to be the dominant contributions, their number on a given time slice is super-polynomial in time. Hence, the applicability of our algorithm in higher dimensions, and whether or not the quantum advantage of the 2D experiment in~\cite{Abanin2025} stands, remains an outstanding open problem.

The plan of the paper is as follows. In Sec.~\ref{sec:setup}, we explain the precise computational problem that our algorithm targets. In Sec.~\ref{sec:key_ideas}, we present an intuitive description of the key physical ideas based on quantum chaos that underlie the algorithm. Sec.~\ref{sec:algorithm} then presents an explicit description of the membrane cluster expansion algorithm for both $\otoc^{(1)}$ and $\otoc^{(2)}$, and a discussion of its runtime and error analysis. Sec.~\ref{sec:numerics} presents numerical results obtained from practical implementations of the algorithm on individual instances of 1D random circuits, and provides strong evidence for the practical efficiency as well as accuracy of the algorithm  for both $\otoc^{(1)}$ and $\otoc^{(2)}$ (see in particular Fig.~\ref{fig:otoc_numerics}, which contains all our main numerical results). We discuss the broader applications of our algorithm and open questions in Sec.~\ref{sec:outlook}.

\section{The computational problem of estimating OTOCs}
\label{sec:setup}

\begin{figure}[!t]
\includegraphics[width=\columnwidth]{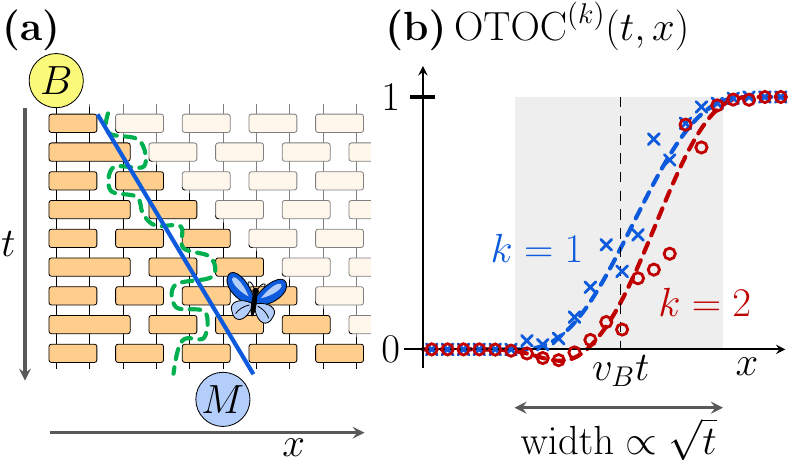}
\caption{{\bf(a)}: Circuit geometry and locations of $B$ and $M$. By light-cone cancellation, only the darker-shaded gates within the causal cone of $B$ contribute to $\otoc^{(k)}$. In the random circuit average, the right endpoint of the support of $B$ follows the ``butterfly cone'' at velocity $v_B$ (blue line). In a single realization, the typical right endpoint fluctuates around this cone (dashed green line). \textbf{(b)}: Schematic profile of the mean values of $\otoc^{(1)}$ (blue dashed) and $\otoc^{(2)}$ (red dashed). The fluctuations around these mean values in individual circuit instances (blue crosses and red circles as  examples)  are restricted to a sensitive regime of $x$ within a width $\propto \sqrt{t}$ of $v_B t$ (gray region).}
\label{fig:circuit_setup}
\end{figure}

We consider a random unitary circuit~\cite{emersonPseudoRandomUnitaryOperators2003,oliveiraGenericEntanglementCan2007,znidaricExactConvergenceTimes2008,fisherRandomQuantumCircuits2023} of depth $t$ in the brickwork geometry of Fig.~\ref{fig:circuit_setup}~(a). All gates are independent Haar-random unitaries on their support. The local Pauli operators $B$ and $M$ in the OTOC are placed at lattice sites $0$ and $x$, respectively, defining the target observable
\begin{equation}
\otoc^{(k)}(t, x) = \frac{1}{2^n}\text{Tr}\left[ \left(B_0(t)M_x\right)^{2k}\right],    
\end{equation}
where $B_0(t) = U(t)^\dagger B_0 U(t)$, $t$ is the circuit depth (sometimes suppressed in $U(t)$ for simplicity), and $n > t + 1$ is the number of qubits. The unitary gates outside the causal cone of $B$ (shown in the lighter shade in Fig.~\ref{fig:circuit_setup}) cancel in the normalized trace.  Hence, only gates within this cone are relevant for the value of $\otoc^{(k)}$, creating an effective $n=t+1$-qubit problem.  

In 1D, $\otoc^{(k)}$ typically exhibits a spacetime profile as shown in Fig.~\ref{fig:circuit_setup}~(b), where the crossover interpolating between $0$ and $1$ occurs at a butterfly front $x \approx v_B t$~\cite{aleinerMicroscopicModelQuantum2016,khemaniVelocityDependentLyapunovExponents2018,xuLocalityQuantumFluctuations2019}, with $v_B=0.6$ denoting the butterfly velocity \cite{nahum_op, vk_op,hunterjonesOperatorGrowthRandom2018}. The OTOC exhibits substantial circuit-to-circuit fluctuations only in the front region of width $\sim \sqrt{t}$ around $x=v_Bt$ (see Refs~\cite{nahum_op, vk_op, mele2026verifiablequantumadvantagerandom} and App.~\ref{app:data_otoc2_fluct}). Due to the sample complexity of estimating Born probabilities, quantum experiments are sample-efficient only within this regime. We refer to this as the {\it sensitive regime}.

The precise task we undertake in this work is to approximate $\otoc^{(k)}$ for $k=1,2$ within the sensitive regime of individual realizations of 1D random circuits to high accuracy at a specified circuit depth $t$. An algorithm is computationally efficient if its runtime scales polynomially in $t$ for the target accuracy, which can for example be inverse polynomial in $t$. (See~\cite{kingSimplifiedVersionQuantum2025} for a similar description of the problem.)

To quantify and compare the accuracy of the approximation, we evaluate two metrics over samples of circuit realizations. The first is the sample mean of the absolute error, which directly assesses closeness to the true value. The second is the signal-to-noise ratio (SNR)~\cite{Abanin2025}, which measures how faithfully the approximation captures circuit-to-circuit fluctuations. The exact and approximate values of $\otoc^{(k)}$ form two random variables over the circuit ensemble. The SNR is defined as $\frac{1}{\sqrt{2(1-\rho)}}$, where $\rho$ is the normalized covariance of the two random variables. Uncorrelated variables give an SNR of $\frac{1}{\sqrt{2}}$, with larger values indicating a stronger correlation.

As a baseline, Google's 2D experiments~\cite{Abanin2025} achieved an estimated 
${\rm SNR} \approx 3.5$ for $\otoc^{(1)}$ at circuit depth $29$, and ${\rm SNR} \approx 2$--$3$ for $\otoc^{(2)}$ at depth $23$.

The circuit geometry in Fig.~\ref{fig:circuit_setup} has a ``flat edge'' where three-qubit boundary gates appear at the left boundary in every second layer. This gate structure does not affect the efficacy of our algorithm, but reduces boundary effects in the numerics; moreover, this geometry is experimentally feasible. We also consider a setup where we place the operator $B$ far  from the left boundary, which is equivalent to the case of an infinite chain as long as the distance from the boundary is larger than $t+1$. This setup increases the practical computational cost of our algorithm.  However, the complexity in this setup remains polynomial in $t$, and we are able to achieve comparable accuracy in this case to the near-boundary case for $\otoc^{(1)}$ (Figs.~\ref{fig:otoc_numerics}(b1)-(c1)).

\section{Physical basis for the algorithm} 
\label{sec:key_ideas}

In this section, we explain the essential physical intuition from quantum many-body chaos that underlies our membrane cluster expansion algorithm, before presenting the concrete implementation of the algorithm in Sec.~\ref{sec:algorithm}.

$\otoc^{(k)}$ can be seen as a transition amplitude between replicated local operators under $2k$ forward copies and $2k$ backward copies of the time evolution, $(U \otimes U^{\ast})^{\otimes 2k}$. In the study of quantum information dynamics in chaotic quantum many-body systems, it has recently been understood that for many physically interesting observables on $2k$ copies, the dominant configurations that survive at late times, and underlie phenomena such as thermalization, are associated with {\it pairings} between forward and backward copies of the time evolution~\cite{zhou_entanglement_2020, eq_approx, mark2024maximum,mcginley2025scrooge,Mok2026Nature, VardhanMoudgalya2026,jonay,jonayPhysicalTheoryTwoStage2024,chanSolutionMinimalModel2018,garrattLocalPairingFeynman2021,hoExactEmergentQuantum2022,baoTheoryPhaseTransition2020,jianMeasurementInducedCriticalityRandom2020}.  For the $k=1$ case associated with $\otoc^{(1)}$, there are two such pairings:  $\wick{\c1 U \otimes \c1 U^* \otimes \c2 U \otimes \c2 U^*}$, which we denote as $+$, and $\wick{\c2 U \otimes \c1 U^* \otimes \c1 U \otimes \c2 U^*}$, denoted as $-$ (see Fig.~\ref{intuition_figure}~(a)). In terms of states in the replicated Hilbert space, these configurations can be seen as tensor products of all $\ket{+}$ or all $\ket{-}$ states defined in eq.~\eqref{pm_def} below. More generally, for higher $k$, the analogous configurations are labeled by permutations  $\sigma \in \sS_{2k}$, pairing the $i$'th forward evolution with the $\sigma(i)$'th backward evolution. The late-time dominance of  such paired configurations can be understood from the fact that their contributions to physical quantities are stationary with time, while other configurations typically carry temporally fluctuating phases, which cancel and dissipate away at sufficiently late times.  

On viewing $\otoc^{(k)}$ as a transition amplitude under $(U \otimes U^{\ast})^{\otimes 2k}$, the boundary condition from the operator $B$ induces a ``domain wall'' between permutations in $\sS_{2k}$. For example, $\mathrm{OTOC}^{(1)}$ induces a boundary condition with homogeneous $+$ pairings everywhere in the system, except at the location with the $B$ operator insertion, where the $-$ permutation is favored. Unlike the configurations with one homogeneous pairing, such ``domain wall'' configurations are no longer stationary. We can in principle have nontrivial dynamics, including excursions outside the subspace of permutation configurations,  in a growing region around the domain wall. However,  due to the strongly fluctuating phases associated with non-permutation configurations, long excursions or ``bubbles'' outside the permutation subspace are exponentially suppressed in their duration~\cite{zhou_entanglement_2020}. This chaotic dephasing effect is the key physical mechanism underlying our algorithm.  
  
With this understanding of permutation domain walls, there is no fundamental difference between the case of $\otoc^{(1)}$ and $\otoc^{(2)}$. In the latter case, the initial condition leads to  configurations which can be seen as composites of up to three ``elementary domain walls'' from the $\otoc^{(1)}$ case. Configurations which continue to stay within this three domain wall subspace have the lowest  suppression in time among all configurations on the 8-copy Hilbert space. The three domain walls can evolve separately (see Fig.~\ref{intuition_figure}~(d)), so that we have a larger number of minimally suppressed  trajectories  to keep track of.  Crucially, the number of possible nodes of these trajectories at any given time slice is still polynomial in time in 1D. Again, there can be local excursions out of this subspace into bubble contributions, but such excursions are exponentially suppressed by the same dephasing mechanism discussed above.

In the next section, we will show that the physical picture above can be formulated as truncated cluster expansion of irreducible weights, which translates into a polynomial-time algorithm for both $\otoc^{(1)}$ and $\otoc^{(2)}$.

\section{The membrane cluster expansion algorithm} 
\label{sec:algorithm}

In this section, we provide an explicit description of our algorithm. We will explain the connection of the formal steps here to the key ideas presented in pictorial language in Sec.~\ref{sec:key_ideas}, but this section can be understood independently of the physical picture.  The problem our algorithm tackles is the following:

\textbf{Input:} Description of a brickwork circuit $U$ of depth $T$ on a 1D lattice (see Fig.~\ref{fig:circuit_setup}), local operators $B_0$ at position $0$ and $M_x$ at location $x$, and a cutoff time interval  $\Delta$. 

\textbf{Output:}  An approximation for $\otoc^{(k)}(T,x)$ for the given circuit, which we call $\otoc^{(k)}_{\Delta}(T,x)$.   

We will focus on the cases $k=1, 2$, and comment briefly on higher $k$. The precise meaning of $\Delta$ will be explained below; intuitively, it is a cutoff on the maximum duration of the ``bubbles'' discussed in the previous section. The runtime of the algorithm is ${\rm poly}(T)e^{O(k\Delta)}$. We will argue that under a formal version of the ``bubble suppression'' assumption described intuitively in Sec.~\ref{sec:key_ideas}, the error $|\otoc^{(k)}-\otoc^{(k)}_{\Delta}|$ is upper-bounded by ${\rm poly}(T)e^{-\Omega(\Delta)}$ in the sensitive regime~(Fig.~\ref{fig:circuit_setup}~(b)).\footnote{$f(x) = O(g(x))$ indicates that $f(x)$ is upper-bounded by some positive constant times $g(x)$, while $f(x) = \Omega(g(x))$ indicates that $f(x)$ is lower-bounded by some positive constant times $g(x)$.} Hence, we expect that any desired inverse polynomial error in the OTOCs can be achieved in polynomial runtime. 

Crucial for our algorithm is the representation that we use for the OTOCs. We therefore first explain this representation, an exact series for $\otoc^{(k)}$ which we call the $d/b$ decomposition, in Sec.~\ref{sec:db}. A key ingredient of this decomposition is the choice of a certain ``simple subspace,'' which we illustrate for $\otoc^{(1)}$ in Sec.~\ref{sec:db}.  We then explain how this series can be truncated to obtain an efficient algorithm for both OTOCs in Sec.~\ref{sec:otoc1},  and estimate the errors incurred by the truncation in Sec.~\ref{sec:algorithm_cost_error_main}. The choice of simple subspace for $\otoc^{(2)}$ involves further physical inputs,  discussed in Sec.~\ref{sec:otoc2}.

\subsection{The $d/b$ decomposition}
\label{sec:db}

We view $\otoc^{(k)}$ as a transition amplitude in the replicated Hilbert space $(\mathcal H\otimes \overline{\mathcal H})^{\otimes 2k}$, consisting of  $2k$ copies of the original Hilbert space $\mathcal H$, and $2k$ copies of the ``conjugate" space $\overline{\mathcal H}$ which is time-evolved by $U^{\ast}$. Explicitly, we have 
\begin{equation}
{\otoc}^{(k)}(T,x)= \frac{1}{2^n}\langle\Psi_{\mathrm{in}}(B_0)|\mathcal{U}_1\cdots\mathcal{U}_T |\Psi_{\mathrm{out}}(M_x)\rangle \label{otoc_1_amp}
\end{equation}
Here, we have labeled the circuit layers from top to bottom by $t=1,\ldots,T$, and defined $\mathcal U_t=(U_t\otimes U_t^*)^{\otimes 2k}$, where $U_t$ is the $t$'th layer. The  precise definitions of  $\ket{\Psi_{\mathrm{in}}(B_0)}$ and $\ket{\Psi_{\mathrm{out}}(M_x)}$  are given in Appendix~\ref{app:higher_point}. We will in general have in mind the evolution from the ``top'' state associated with $B$  to the ``bottom'' state associated with $M$.

\textbf{The simple subspace:} The $d/b$ decomposition divides the $4k$-copy Hilbert space at each intermediate time $t$ into a {\it simple subspace} and its complement. A good choice of simple subspace is one that contains a small number of states (polynomial in $t$), yet those states ``dominate'' the contributions to $\otoc^{(k)}$ in a sense we will make precise below. 

Let us first illustrate the choice of simple subspace for $\otoc^{(1)}$. Although our goal is to compute
$\otoc^{(1)}(T,x)$ for individual circuit realizations, the ensemble average $\mathbb{E}_U[\otoc^{(1)}(T,x)]$ provides a useful guide to choosing the simple subspace. Under Haar-averaged gates, the evolution of  $\bra{B_0, +}$ lies purely within the span of states that are tensor products of the following two {\it permutation states} on 4 copies of a single qubit, $\sH_1 \otimes \bar \sH_{1'} \otimes \sH_2 \otimes \bar \sH_{2'}$~\cite{collinsIntegrationRespectHaar2006}: 
\begin{equation}
|+\rangle=|\Phi\rangle_{1,1'}|\Phi\rangle_{2,2'},\quad
|-\rangle=|\Phi\rangle_{1,2'}|\Phi\rangle_{2,1'}. \label{pm_def} 
\end{equation}
Here, $\ket{\Phi}_{a,b} =|0\rangle_a|0\rangle_b+|1\rangle_a|1\rangle_b$ is an unnormalized maximally entangled state. These two states precisely refer to the $+$ and $-$ pairings between the copies discussed in Sec.~\ref{sec:key_ideas} and Fig.~\ref{intuition_figure}~(a). In fact, the evolution of $\bra{B_0, +}$ within the random average only involves configurations with at most a {\it single domain wall} between these $\ket{+}$ and $\ket{-}$ states on different sites. 
More explicitly, at time slice $t$ below the unitary layer $U_t$, lightcone cancellation restricts the relevant dynamics to $(\mathcal H_t\otimes\overline{\mathcal H_t})^{\otimes 2}$, where $\mathcal H_t$ is the Hilbert space of $t+1$ qubits starting from the left edge. Within this space, the relevant states with up to one domain wall are given by 
\begin{equation}
\ket{D_x}
=
\ket{-}\cdots\ket{-}\textcolor{softmaroon}{{\Big|}_{x}}
\ket{+}\cdots\ket{+},
\qquad x\in\mathcal X_t.
\label{dx}
\end{equation}
The allowed positions $\mathcal{X}_t$ of the domain wall (indicated in color above) are between complete gate blocks in $U_t$, including the cases to the left of the left-most gate (which has no domain wall) and the right of the right-most gate.

Now in a single instance of the circuit, the evolution of $\bra{B_0, +}$ is no longer restricted to live within this domain wall subspace. However, the key insight underlying our algorithm is that this subspace nevertheless gives the dominant contribution to $\otoc^{(1)}$. We therefore choose the domain wall subspace as our {\it simple subspace} for the simulation of $\otoc^{(1)}$. For $\otoc^{(2)}$, the evolution takes place in a larger Hilbert space, $(\mathcal H_t\otimes\overline{\mathcal H_t})^{\otimes 4}$. We analogously identify a simple subspace $\sS_t$ within this space.  Instead of  just single domain-wall states, $\sS_t$ has states with a general label $\ket{\vec{s}}$ which  parameterizes a space of up to three domain walls of various types. Importantly, $\vec s$  still takes only $O({\rm poly}(t))$ possible values. This subspace will be explained in more detail in Sec.~\ref{sec:otoc2}. The rest of this subsection describes the $d$/$b$ decomposition  for both $\otoc^{(1)}$ and $\otoc^{(2)}$, and we use $\sS_t, \ket{\vec s}$ to jointly denote the simple subspace and simple states for both cases.

\textbf{The $\otoc$ as a sum over histories:} The key element of the $d/b$ decomposition is simply a resolution of the identity in terms  of  $\hat d_t$, the projector onto the simple subspace at time $t$, and the projector onto its orthogonal subspace. For example, for $\otoc^{(1)}$,  $\hd_t
=\sum_{x\in\mathcal X_t}
\ket{D_x^*}\bra{D_x}$, where $\ket{D_x^*}$ refers to the   ``dual state'' of  $\ket{D_x}$, satisfying
$\langle D_x|D_y^*\rangle=\delta_{xy}$ (an explicit construction can be found in Appendix~\ref{app:duals}; the need to define such duals arises from the non-orthogonality of $\ket{+}$ and $\ket{-}$).\footnote{
Technically, $\hd_t$ is an oblique rather than an orthogonal
projector: it satisfies $\hd_t^2=\hd_t$, but generally
$\hd_t^\dagger\neq\hd_t$.
} On inserting this resolution of the identity below $\mathcal{U}_t$, we find from the unitarity of each gate  that $\mathcal{U}_t d_t = d_t$.  This will also be a feature of the simple subspace for $k=2$. We have the exact identity 
\begin{equation}
    \mathcal{U}_t = \hd_t + \hb_t 
\end{equation}
where $\hb_t=\mathcal U_t(\id-\hd_t)$ depends on the specific circuit realization, and $\hat d_t$ is the same for each realization. We have used the intuitive notation $d$ for ``domain wall,'' and $b$ for ``bubble,'' as the latter contribution precisely captures excursions outside the simple subspace. 

For a given circuit, the OTOCs can therefore be written as the following sum: 
\begin{align}
&{\otoc}^{(k)}(T,x)
= \frac{1}{2^n}\inst\prod_{i=1}^{T}(\hat d_i+\hat b_i)\out \nonumber\\
& =\frac{1}{2^n}\sum_{\hat p_i\in \{\hd_i,\hb_i\}}\inst\hat p_1\ldots \hat p_{T}\out.  \label{eq:expansion}
\end{align}
That is, we can explicitly decompose the amplitude into {\em histories}, described by the string $\mathbf{p} \in \{d,b\}^T$. The projector $p_i=d$ if, at time slice $i$, the evolution stays within states in the simple subspace; $p_i=b$ if the evolution goes outside this subspace. Consecutive $b$'s describe an interval during which the evolution never enters the simple subspace; we think of this visually as a ``bubble" (see \Cref{intuition_figure}).

As a step towards developing the algorithm, it is useful to consider the following closely related quantity: the transition amplitude between two simple states, $\ket{\vec{s_1}}$ at time slice $t_1$, and $\ket{\vec{s_2}}$ at  $t_2>t_1$. This quantity captures contributions to \eqref{eq:expansion} coming from layers between the top and bottom layers, and is defined as 
\begin{align}\label{eq:Z}
Z(t_1,\vec s_1;t_2,\vec s_2)
&=\bra{\vec s_1 }\mathcal U_{t_1 + 1}\cdots\mathcal U_{t_2 }
\ket{\vec s_2^{~\ast}}
\end{align}
where the dual state $\ket{{\vec s_2}^{~\ast}}$ is defined by $\braket{\vec s|\vec{s_2}^{\ast}}= \delta_{\vec s_2, \vec s}$ for any $\vec s$. Writing each $\mathcal{U}_i = \hat d_i + \hat b_i$, this amplitude  again takes the form of a sum over histories, and is thus denoted as a partition function $Z$.  For $t_2> t_1+1$,  we single out the term in $Z$ with no intermediate $\hd$ insertions, and denote it by $W$: 
\begin{equation}
\begin{aligned}
W(t_1,\vec s_1;t_2,\vec s_2)
&=\bra{\vec s_1 }\hb_{t_1+1}\cdots\hb_{t_2-1}
\ket{\vec s_2^{~\ast}}.
\end{aligned}
\label{eq:irreducible_amplitude}
\end{equation}
$W$ can be seen as an {\it irreducible amplitude}, or a bubble which cannot be decomposed into smaller bubbles.  For $t_2 = t_1+1$, the product of $\hb$'s is empty, so we use the convention $W(t_1,\vec s_1;t_1+1,\vec s_2)
=Z(t_1,\vec s_1;t_1+1,\vec s_2)$. In this case, $W$ should be seen as a one-step propagator, rather than a bubble.  

\textbf{A recursion relation for the transition amplitude:} In the decomposition of $Z$ as a sum over histories, by grouping all $2^{t_2-t_1}$ histories by the first time from the top boundary where they return to the simple subspace, and inserting the projector $\hat d_t$ at that time slice, we obtain the following recursion relation:
\begin{align}
&Z(t_1,\vec s_1;t_2,\vec s_2)
=W(t_1,\vec s_1;t_2,\vec s_2) \nonumber\\
&\quad+\sum_{t'=t_1+1}^{t_2-1}\sum_{\vec s\,'\in\mathcal \sS_{t'}}
W(t_1,\vec s_1;t',\vec s\,')Z(t',\vec s\,  ';t_2,\vec s_2).\label{eq:Z_recursion}
\end{align}
For example, in the case of $\otoc_1$, this recursion relation can be simply represented as a spacetime diagram, where time decreases downwards, and a vertex indicates the position of a domain wall:
\begin{equation}
\label{rec2_diagram}
\begin{tikzpicture}[baseline={(0,0.7)},x=1cm,y=1cm]
\draw[thick,double distance=1.5pt] (0.45,0) -- (0,1.4);
\node[left] at (0.225,0.7) {$Z$};
\fill (0.45,0) circle (1.3pt)
    node[below] {\scriptsize $(t_2,x_2)$};
\fill (0,1.4) circle (1.3pt)
    node[above] {\scriptsize $(t_1,x_1)$};

\node at (0.90,0.7) {$=$};

\begin{scope}[xshift=1.55cm]
\draw[thick,fill=blue!10]
    (0.45,0) to[out=147.82,in=247.82] (0,1.4)
    to[out=-32.18,in=67.82] (0.45,0);
\node at (0.225,0.7) {$W$};
\fill (0.45,0) circle (1.3pt)
    node[below] {\scriptsize $(t_2,x_2)$};
\fill (0,1.4) circle (1.3pt)
    node[above] {\scriptsize $(t_1,x_1)$};
\end{scope}

\node at (2.60,0.7) {$+$};
\node at (3.65,0.7)
    {$\displaystyle
      \sum_{t'=t_1+1}^{t_2-1}\;
      \sum_{x'\in\mathcal{X}_t}$};

\begin{scope}[xshift=5.15cm]
\draw[thick,double distance=1.5pt] (0.45,0) -- (0.06,0.6);
\node[left] at (0.255,0.3) {$Z$};

\draw[thick,fill=blue!10]
    (0.06,0.6) .. controls (-0.518,0.813) and (-0.442,1.133) ..
    (-0.4,1.4) .. controls (0.178,1.187) and (0.202,0.867) ..
    (0.06,0.6);
\node at (-0.15,1.0) {$W$};

\fill (0.45,0) circle (1.3pt)
    node[below] {\scriptsize $(t_2,x_2)$};
\fill (0.06,0.6) circle (1.3pt)
    node[right=2pt] {\scriptsize $(t',x')$};
\fill (-0.4,1.4) circle (1.3pt)
    node[above] {\scriptsize $(t_1,x_1)$};
\end{scope}
\end{tikzpicture}
\end{equation}
Here a double line represents the sum over all histories
contributing to $Z$, and a shaded bubble represents an
irreducible amplitude $W$. Running the recursion shows that $Z$ can  be written as sum of all possible  products of $W's$, i.e., all possible diagrams like the one shown in Fig.~\ref{intuition_figure}~(b) (so far with no truncation). The irreducible weights $W$ are thus ``linked clusters," which can be chained together to produce the full expansion. Our algorithm is named based on this analogy.  

\subsection{The membrane cluster expansion algorithm }
\label{sec:otoc1}
Our algorithm can now be summarized in one sentence: it approximates the exact recursion relation \eqref{eq:Z_recursion} for $Z$ by keeping only the $W$'s of temporal duration at most $\Delta$; that is, it computes the quantity in eq. \eqref{eq:Z_Delta_recursion} below. This gives the ``bulk contribution'' to $\otoc^{(k)}$ away from the top and bottom boundaries. To get an approximation for $\otoc^{(k)}$ itself, the boundaries can then be glued to this bulk contribution in a precise way described at the end of the section. 

Formally, we define the truncated irreducible amplitude, where the only retained cases are of duration $\leq \Delta$:
\begin{equation}
W_{\Delta}(t_1,\vec s_1;t_2,\vec s_2)=
\begin{cases}
W(t_1,\vec s_1;t_2,\vec s_2),&1\leq t_2-t_1\leq\Delta,\\
0,&\text{otherwise}.
\end{cases}
\label{eq:W_truncated}
\end{equation}
As explained below, $W_{\Delta}$ can be computed with cost roughly exponential in $\Delta$. Our algorithm then outputs the approximation $Z_{\Delta}$, recursively defined in terms of the $W_{\Delta}$'s as follows:  
\begin{align}
&Z_{\Delta}(t_1,\vec s_1;t_2,\vec s_2)
=W_{\Delta}(t_1,\vec s_1;t_2,\vec s_2)\nonumber\\
&+\sum_{t'=t_1+1}^{t_2-1}\sum_{\vec s\,' \in\mathcal \sS_{t'}}
W_{\Delta}(t_1, \vec s_1;t',\vec s\,')Z_{\Delta}(t',\vec s\,';t_2, \vec s_2). \label{eq:Z_Delta_recursion}
\end{align}
Physically, we expect this approximation to be valid due to the strong suppression of large bubbles by  chaotic dephasing, which was discussed heuristically  in Sec.~\ref{sec:key_ideas}. In Sec.~\ref{sec:algorithm_cost_error_main} and Appendix~\ref{app:total_error_bound}, we will present this physical intuition more formally as a ``bubble suppression" condition, and discuss how the accuracy of the approximation follows from it. 

The algorithm to compute $Z_{\Delta}$ has two computational steps:

\textbf{Step 1: compute all short $W$'s} --- We compute all $W$'s of short duration which are needed as input for the recursion~\eqref{eq:Z_Delta_recursion}. The computation of $W$ could in principle be carried out by an exact  contraction of the  tensor network in its definition~\eqref{eq:irreducible_amplitude}. However, practically, the computation of even short $W$'s is expensive due to the nonlocal projectors $(\mathbb I - \hd_t)$ inserted at every intermediate time slice. The exact $Z$ amplitudes for short durations are easier to evaluate, and we derive $W$ from them by inverting the exact recursion relation~\eqref{eq:Z_recursion}. The computational cost scales as $e^{O(k\Delta)}$, since computing $Z$ requires a tensor contraction whose widest cross section contains $4k$ copies of $O(\Delta)$ qubits.  The cost is thus much larger in practice for the case of $\otoc^{(2)}$ compared to $\otoc^{(1)}$.

\textbf{Step 2: sum the truncated expansion by dynamic programming} ---
Having obtained all $W_{\Delta}$, the remaining step is simply to sum the recursion relation~\eqref{eq:Z_Delta_recursion}. A naive evaluation would enumerate exponentially many histories. Instead, we use \emph{dynamic programming}: we compute $Z_\Delta$ for progressively longer time
intervals, storing and reusing the intermediate computations of $Z_{\Delta}$ for shorter intervals. Crucially, in one spatial dimension, the number of simple subspace states for both $\otoc^{(1)}$ and $\otoc^{(2)}$ at any time slice  at most polynomial in the total time $T$. Hence, the number of intermediate amplitudes that must be stored and combined is polynomial, even though the recursion sums exponentially many histories.

Together, the two steps for computing $Z_{\Delta}$ have a cost upper-bounded by ${\rm poly}(T) e^{O(k\Delta)}$. We will now explain the slightly modified treatment of the boundary conditions needed to evaluate $\otoc^{(k)}_{\Delta}$ instead of $Z_{\Delta}$, which leaves the scaling of this cost unchanged.

For the OTOCs, the physical boundary states associated with $B_0$ and $M_x$ do not lie within the simple subspace. Here, we consider histories which  reach the simple subspace within some window at both ends of the circuit. In particular, we consider histories with at least one $\hd$ insertion in each of the windows $1,\ldots,\Delta_{\rm top}$ and $T-\Delta_{\rm bottom},\ldots,T-1$, for some chosen $\Delta_{\rm top}$ and $\Delta_{\rm bottom}$, both no smaller than $\Delta$. 
For each such history, let $j$ be its last $\hd$ insertion in the top window and $\ell$ its first $\hd$ insertion in the bottom window. All histories above $j$ and below $\ell$ are summed exactly, giving the boundary amplitudes
\begin{equation}
    \begin{aligned}
        &Z_{\rm top}(j,\vec s_j)
 = \inst \mathcal U_1\cdots\mathcal U_{j-1}\ket{\vec s_j^{\,*}},\\
 & Z_{\rm bottom}(\ell,\vec s_\ell)
 = \bra{\vec s_\ell}\mathcal U_{\ell+1}\cdots\mathcal U_T\out.
    \end{aligned}
\end{equation}
Between $j$ and $\ell$, successive $\hd$ insertions are connected by the truncated bulk irreducible weights $W_{\Delta}$ of eq.~\eqref{eq:W_truncated}. We then obtain the  approximation  $\otoc^{(k)}_\Delta(T,x)$ by summing the product of the two boundary amplitudes and the bulk $W$'s over the allowed insertion times and simple-state labels  (with the normalization $2^{-n}$ of Eq.~\eqref{eq:expansion}). In the flat-edge geometry of Fig.~\ref{fig:circuit_setup}, we take $\Delta_{\rm bottom}=\Delta$ and allow $\Delta_{\rm top}\geq\Delta$ to capture longer excursions near $B_0$. See Appendix~\ref{app:boundary_layer} for  more details.

It is worth noting that for $\otoc^{(1)}$, the lower boundary condition  $\ket{\Psi_{\rm out}(M_x)}$ consists of  $\ket{-}$ states at all  sites other than $x$, which favors the expansion of $\ket{-}$ domains in going from the top to the bottom, at velocity $v_B=0.6$ (see Fig.~\ref{fig:circuit_setup}). Hence, even though our algorithm sums over all possible histories up to the restriction on bubble duration, the dominant contributions to the sum over histories in \eqref{eq:expansion} come from cases where domain walls at most intermediate steps  are close to a path with velocity $v_B$. Similar paths are expected to dominate contributions to $\otoc^{(2)}$.

When $B_0$ is far from the boundary, we use a bulk expansion of the same form~\eqref{eq:expansion}, with a modified choice of simple subspace. For $\otoc^{(1)}$,  states in the simple subspace are now labelled by the positions $\vec s = (x_1, x_2)$ of  two domain walls that bound a single $\ket{-}$ domain in a  background of $\ket{+}$ states (see Appendix~\ref{app:two_dw}).

\subsection{Bounds on error and computational cost}
\label{sec:algorithm_cost_error_main}

\begin{figure}[!t]
\centering
\includegraphics[width=\columnwidth]{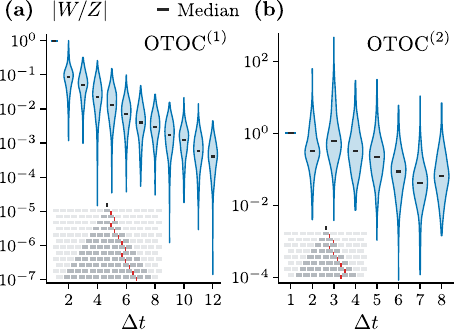}
\caption{Relative bubble weights $|W/Z|$ for propagation between few-wall states in (a) $\otoc^{(1)}$ (400 circuit samples) and (b) $\otoc^{(2)}$ (200 circuit samples). Both $W$ and $Z$ start at the  black bar and end at one of the red bars after $\Delta t$ completed layers.  The domain wall locations of the  endpoints (red bars) are chosen to be nearest to $x=v_B\Delta t$, with $v_B=0.6$. The horizontal width of each violin is proportional to the estimated probability density of $\log_{10}|W/Z|$, rescaled to a common maximum width. Horizontal black bars indicate medians.  Both plots show overall exponential decay of $|W/Z|$ with $\Delta$ at sufficiently large $\Delta$, with the increase between the second-to-last and last point in (b) arising from a lattice effect explained further in the text (see Sec.~\ref{sec:numerics}).  }
\label{fig:w_over_z}
\end{figure}

The key physical input from quantum chaos that ensures the fast convergence of the membrane cluster expansion is the typical relative exponential decay
\begin{equation}
    \left|\frac{W(t_1, \vec s_1; t_2, \vec s_2)}{Z(t_1, \vec s_1; t_2, \vec s_2)}\right| \leq C e^{-\lambda(t_2 - t_1)} \label{w_z_decay}
\end{equation}
for some positive constants $C$ and $\lambda$. We refer to this as the {\it bubble suppression condition}. Among different circuit realizations, we expect this assumption to fail only with exponentially small probability in  $\Delta t = t_2 - t_1$ when $\Delta t$ is above some constant value.  In Fig.~\ref{fig:w_over_z}, we demonstrate this exponential decay numerically for the $W$ and $Z$ relevant for  $\otoc^{(1)}$ and $\otoc^{(2)}$. We consider velocities $\Delta x/\Delta t$ close to $v_B$, where $\Delta x$ is the spatial distance between the domain walls at the endpoints of $W$ and $Z$. As discussed at the end of Sec.~\ref{sec:otoc1}, such velocities are expected to give the dominant contributions to the OTOCs. We provide further evidence for this condition in random circuits in Appendix~\ref{app:WZ_decay_argument}, including a theoretical argument in large $q$ limit, and explicit calculations about the ``failure probability" in the case of a vertical domain wall which is shown to decay exponentially in time. Since this condition captures an intuition about chaotic dephasing which applies much more broadly, we also expect it hold in other sufficiently chaotic circuit families beyond Haar-random circuits.   

Under the above bubble suppression condition, as well as a mild convexity condition which captures the idea that non-straight trajectories give more suppressed contributions than straight ones, we argue that the relative error in $Z_{\Delta}$ for a time interval $T$ obeys the bound 
\begin{equation}\label{total_error}
\left|\frac{Z-Z_\Delta}{Z}\right|\leq \text{poly}(T) e^{-\lambda\Delta}.
\end{equation}
The formal argument for this bound is presented in Appendix~\ref{app:total_error_bound}, and we provide some physical intuition for it at the end of this subsection. We further expect that for the OTOCs in the sensitive regime, this bound for $Z$  translates to an additive error $|\otoc^{(k)}(T, x) - \otoc^{(k)}_{\Delta}(T, x)| \leq {\rm poly}(T) e^{-\lambda \Delta}$. Hence, for a target error $\epsilon$, the required $\Delta$ scales as $O(\log T + \log (1/\epsilon))$. 
The circuit-to-circuit fluctuations in $\otoc^{(k)}$ in the sensitive regime are $\frac{1}{{\rm poly}(T)}$,\footnote{This has been shown for $k=1$~\cite{nahum_op, vk_op, mele2026verifiablequantumadvantagerandom}, and is expected to also be true for $k\geq 1$.} so that it is natural to choose an error threshold $\epsilon= \frac{1}{{\rm poly}(T)}$. Combining this with the understanding from the previous section that the runtime of implementing the algorithm is ${\rm poly} (T)e^{O(k\Delta)}$, we see that any inverse polynomial error can be achieved in polynomial runtime. The power of $T$ in this polynomial runtime is higher for the $k=2$ case compared to $k=1$ for a given threshold.  

To understand intuitively how the error~\eqref{total_error} accumulates over all the truncations, it is useful to think in the pictorial language of domain walls and bubbles in Fig.~\ref{intuition_figure}(b). The bubble suppression condition ensures that by truncating the series, the largest bubbles that we leave out of the exact sum contribute a relative error  of order $e^{-\lambda \Delta}$. There are $O(T)$ choices of starting times where these bubbles can be inserted, and another ${\rm poly}(T)$ choices of the input and output states $\vec s$, which naturally gives rise to the estimate  \eqref{total_error} for the total relative error.   

\subsection{The simple subspace for $\otoc^{(2)}$}
\label{sec:otoc2}

In this section, we explain the simple subspace used in our algorithm for $\otoc^{(2)}$. Generalizing the $\ket{+}$ and $\ket{-}$ states of \eqref{pm_def}, in this case there are 24  permutation states $\ket{\sigma}$ for   $\sigma \in \sS_4$ associated with pairings. These states  are defined on 4 copies $a =1, ..., 4$ of a qubit in the original Hilbert space, and 4 copies $a' =1, ..., 4$ in the conjugate space. They correspond to pairing copy $a$ with $\sigma(a)'$, i.e.,  
$\ket{\sigma} = \otimes_{a=1}^4 \ket{\Phi}_{a,\,  \sigma(a)'}$. 

The average of the replicated time evolution operator over Haar random circuits again involves only states within this permutation subspace on each site. However, an important difference between $\otoc^{(1)}$ and $\otoc^{(2)}$ is that the exact evolution of $\inst$ under the averaged time evolution on 8 copies is no longer restricted to a subspace with an $O(1)$ number of domain walls. Instead, the number of states that can appear in the random average at time $t$ grows exponentially with time, violating the required conditions for a simple subspace for our algorithm. 

Nevertheless, a closer look at the random average naturally selects the following choice of simple subspace within $(\mathcal H_t\otimes\overline{\mathcal H_t})^{\otimes 4}$. It is the span of states of the following form, with up to three domain walls (see Fig.~\ref{intuition_figure}(d)): 
\begin{align}
&\ket{(x_1, x_2, x_3), (\sigma_1, \sigma_2)}= \nonumber\\  &\ket{\eta} ... \ket{\eta} \textcolor{softmaroon}{{\Big|}_{x_3}}\ket{\sigma_{12}}...\ket{\sigma_{12}} \textcolor{softmaroon}{{\Big|}_{x_2}}\ket{\sigma_1}...\ket{\sigma_{1}} \textcolor{softmaroon}{{\Big|}_{x_1}} \ket{e}... \ket{e} \label{3dw_states}
\end{align}
Here, $e$ refers to the identity permutation, and $\eta=(1234)$ is a cyclic permutation. $\sigma_{12}= \sigma_2 \circ \sigma_1$, and the trajectory $\eta \to \sigma_{12} \to \sigma_1 \to e$ must satisfy the condition that it lies along a geodesic from $\eta$ to $e$ with respect to the Cayley distance. The Cayley distance is the  minimal number of transpositions between two permutations, or intuitively number of elementary domain walls between them~\cite{hunterjonesUnitaryDesignsStatistical2019}. Again, at time $t$, the allowed positions of the domain walls $x_i$  are between the gates in the layer $U_t$.  Any of the domains is allowed to be empty, but the remaining domains must still satisfy the geodesic condition. The number of states in this subspace is $O(t^3)$. 

While the above subspace appears to be somewhat formal, one can intuitively argue that it is the natural choice of simple subspace based on three observations: 
\begin{enumerate}
\item[(i)] The top boundary condition $\inst$ has $\bra{e}$ states on all sites other than 0. \item[(ii)] The bottom  boundary condition $\out$ has $\ket{\eta}$ states on all sites other than $x$. \item[(iii)] In the large $q$ limit of the average over random  circuits, where $q$ is the on-site Hilbert space dimension,  a domain wall with $\sigma_{\ell}$ to the left and $\sigma_r$ to the right extending for time $t$ gives a bulk contribution proportional to $q^{-d(\sigma_{\ell}, \sigma_r)t}$, where $d$ is the Cayley distance~\cite{zhouEmergentStatisticalMechanics2019}.
\end{enumerate} 
The competing effects of the top and bottom boundaries naturally induce domain wall configurations with  $\eta$ in the leftmost domain, and $e$ in the rightmost domain. The suppression with the Cayley distance in (iii) favors minimizing the total Cayley distance in going from $\eta$ to $e$ among all possible intermediate configurations, which is precisely equivalent to the geodesic condition. These can be seen as the states with up to three ``elementary'' domain walls~\cite{zhouEmergentStatisticalMechanics2019, fmi}.

Hence, for $\otoc^{(2)}$, certain kinds of bubble contributions are present even {\it within} the average over Haar-random circuits (see the top bubble with five domain walls in Fig.~\ref{intuition_figure}(d) as an example).  The membrane cluster expansion can thus be applied even within the random average, to obtain an increasingly accurate estimate of $\mathbb E_U[\otoc^{(2)}]$ by increasing $\Delta$. 
A single random circuit instance has additional bubble contributions beyond those in the random average. All bubble contributions are collectively included in $W$, by taking the simple subspace $\sS_t$ to consist of \eqref{3dw_states}, defining $\hat d_t$ to be the projector onto this subspace, and applying the formalism of Sec.~\ref{sec:db}-\ref{sec:algorithm_cost_error_main}. See Appendix~\ref{app:otoc2} for further details.

In the practical implementation of our algorithm for $\otoc^{(2)}$, due to the computational cost, we consider only the case where $B$ is at the edge of the system. However, the formulation of the polynomial-time algorithm can easily be generalized to the case with operator $B$ in the middle, by including up to 6 domain walls in the simple subspace. Moreover, there is a natural generalization of the geodesic condition that we used to define the simple subspace for higher $k$, which generalizes our polynomial-time algorithm to any finite $k$ with increasing practical computational cost.

\section{Numerical demonstration of efficiency and accuracy}
\label{sec:numerics}

\begin{figure*}[!t]
\centering
\includegraphics[width=\textwidth]{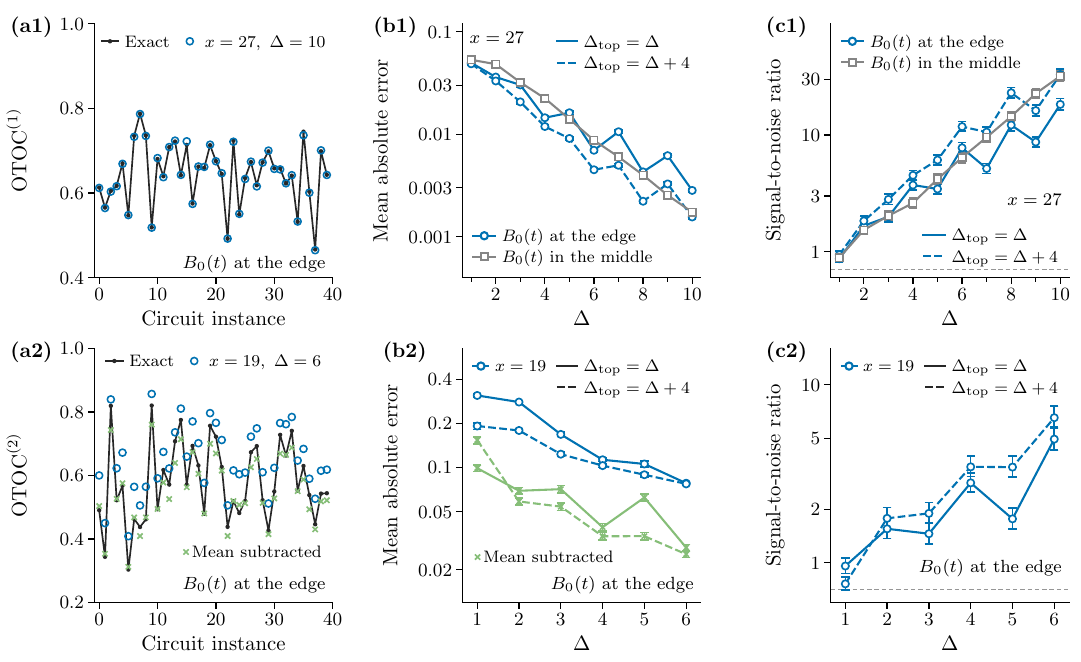}
\caption{Numerical accuracy for {\bf (top)} $\otoc^{(1)}$ at $x = 27, T = 40$ (400 samples) and {\bf (bottom)} $\otoc^{(2)}$ at $x = 19, T = 24$ (200 samples). Both are chosen around the sensitive regime $x \sim v_B T$ with $v_B = 0.6$. {\bf (a1)-(a2)} Comparison of the exact and approximate values for the first 40 circuit instances. {\bf(b1)-(b2)} Mean absolute errors and {\bf (c1)-(c2)} SNR. The vertical axis is in logarithmic scale. Solid and dashed lines distinguish cases with a uniform choice of cutoff $\Delta$ (solid), and a larger cutoff $\Delta_{\rm top}  = \Delta + 4 $ for top boundary layer (dashed). A larger $\Delta_{\rm top}$ mitigates  boundary effects in the semi-infinite chain. (b1)-(c1) also show the comparable accuracy of the $\otoc^{(1)}$ estimate in the infinite chain. Gray dashed lines in (c1)-(c2) mark the  reference value $1/\sqrt{2}$ for uncorrelated random variables.  Data shown with green crosses in (a2)-(b2) is obtained by calibrating the sample mean of our approximation for $\otoc^{(2)}$ with the theoretical mean over random circuits computed by an MPS method (see the text for details). }
\label{fig:otoc_numerics}
\end{figure*}

In this section, we show our results for the practical implementation of the membrane cluster expansion algorithm for $\otoc^{(1)}$ and $\otoc^{(2)}$.

As discussed in Sec.~\ref{sec:algorithm}, the key requirement needed to guarantee the theoretical efficiency of the algorithm is the bubble suppression condition~\eqref{w_z_decay}. Fig.~\ref{fig:w_over_z} provides numerical evidence that this condition is satisfied. The ``violins'' depict the estimated probability density of $\log_{10}|W/Z|$, and the inset circuit shows the starting point (black bond) and endpoints (red bonds) of $W$ and $Z$. The red endpoint bonds are chosen so that the slope of the path is closest to $v_B = 0.6$.  For the single $- \rightarrow +$ domain wall considered in $\otoc^{(1)}$, Fig.~\ref{fig:w_over_z}(a) shows a clear exponential decay for the {\it distributions}. Thus, the bubble suppression condition holds true with overwhelming probability in large times. 
 Here, to capture $|W/Z|$ decay in the bulk of the circuit, we use the ``pyramid'' geometry equivalent to an infinite system at a given time. 

For $\otoc^{(2)}$, we require \eqref{w_z_decay} to be satisfied for all possible choices of $\vec s_1$, $\vec s_2$ in the simple subspace. We perform a partial check of this assumption in Fig.~\ref{fig:w_over_z}(b), by considering the subset of cases where the start and end points of $W$ and $Z$ have a single domain wall between $\eta$ and $e$ (i.e. the middle two domains in \eqref{3dw_states} are empty), and still taking the the velocity to be $v_B$, which is expected to give the dominant contributions. Fig.~\ref{fig:w_over_z}(b) shows a pattern consistent with an exponential decay of $|W/Z|$ at sufficiently large $\Delta t$. The bubble with the largest duration exhibits an apparent deviation from this trend, mainly due to a lattice effect. The red bond at $\Delta t = 8$ experiences a periodic shift to keep the path slope as close as possible to $v_B=0.6$. Nevertheless, the relative weight at $\Delta t = 8$ is nearly an order of magnitude smaller than at $\Delta t = 3$, providing good evidence that the bubble suppression assumption continues to hold at larger $\Delta t$. 

To provide strong evidence for practical as well as theoretical efficiency, we next explicitly implement the  algorithm and show the resulting accuracy in 
Fig.~\ref{fig:otoc_numerics}  for $\otoc^{(1)}$ (top panel) and $\otoc^{(2)}$ (bottom panel). 

In the top panel of Fig.~\ref{fig:otoc_numerics}, (a1) visually demonstrates the closeness of the computed $\otoc^{(1)}$ to the true values in the first 40 samples for the flat edge geometry (Fig.~\ref{fig:circuit_setup}) at $x = 27$ and $T = 40$. The excellent performance is quantitatively exhibited by the exponential decay of the mean absolute error in (b1) and exponential growth of SNR in (c1) for both the flat edge ($B_0$ at left edge), and the effectively infinite chain geometry where $B_0$ is far away from the left edge. 

To mitigate the slightly worse performance and odd-even oscillations caused mostly by the top of the left edge case, we increase the cutoff up to $\Delta_{\rm top}$ in the region close to the top boundary (see Appendix~\ref{app:top} for details). With minor additional computational cost, we boost the accuracy to match or surpass that of the infinite chain. 

The computation of $\otoc^{(2)}$ is a more challenging practical test of the membrane cluster expansion, as the leading three-wall sector contains $O(t^3)$ configurations. Nevertheless, the larger $\Delta_{\mathrm{top}} = \Delta + 4$ choice still allows us to observe an empirical exponential decay of the error and growth of the $\mathrm{SNR}$ for bulk cutoffs up to $\Delta = 6$.   The observed error in Fig.~\ref{fig:otoc_numerics}(b2) still decays with larger $\Delta$, supporting the three-wall truncation. The growth of the $\mathrm{SNR}$ in Fig.~\ref{fig:otoc_numerics}(c2) up to about 6, a result more favorable than the experimental SNR in 2D $\otoc^{(2)}$ at comparable depths~\cite{Abanin2025}, further shows that the algorithm can track circuit-to-circuit fluctuations.

In the comparison with the exact values in Fig.~\ref{fig:otoc_numerics}(a2), our approximation is close to but generally biased above the exact results. This is explained by the fact that  restricting the permutation sector to three domain walls omits contributions even to the ensemble average, introducing an error in the mean (see Sec.~\ref{sec:otoc2}). We thus recalibrated our data as follows: for each $\Delta$, we add a constant to the data points independent of the circuit realization, such that the sample mean of the shifted $\otoc^{(2)}_{\rm \Delta}$ matches the mean $\mathbb E_U [\otoc^{(2)}]$.  We compute the latter using a variational MPS method \cite{Schollwoeck_2011}: using an analytic expression for the action of the averaged gates, we contract the resulting tensor network row by row, fitting the boundary MPS corresponding to a given time site-by-site using alternating least-squares. The results of our computation at $\chi=96$ agreed with those at $\chi=128,160$ to within $10^{-5}$ in absolute value. This calibrated data is shown with the light green crosses in Fig.~\ref{fig:otoc_numerics}(a2)-(b2). The absolute error of the calibrated data thus only captures the errors in  the fluctuations about the mean.  In Fig.~\ref{fig:otoc_numerics}(b2), these absolute errors decay about $4-5$ fold from $\Delta = 1$ to $\Delta = 6$, and seem to show an expected  inverted profile when  compared with the other error measure of fluctuation, the SNR in Fig.~\ref{fig:otoc_numerics}(c2). This check further corroborates the accuracy of the algorithm. 

In Appendix~\ref{app:more_results}, we show additional data on the performance of our algorithm for both $\otoc^{(1)}$ and $\otoc^{(2)}$ for other values of $x$ in the sensitive regime, finding a similar dependence on $\Delta$. 

\textbf{Computational resources.} We briefly discuss the implementation of the algorithm and the computational resources used. The implementation is divided into two stages: computation of all required exact $Z$ values up to duration $\Delta$, and postprocessing of these $Z$ values to obtain the exact short $W$ weights as well as the $\otoc$ observables. For the first stage, we formulate each desired $Z$ value as a tensor network, and contract it on GPUs using complex64 precision. These values are then stored in complex128 and sent to the second stage for observable computation on CPUs. The runtime is dominated by the first stage. For the system size shown in Fig.~\ref{fig:otoc_numerics}, our current implementation takes $\sim 3$ minutes for $\otoc^{(1)}$ with $B$ at the edge, $\sim 1$ hour for $\otoc^{(1)}$ with $B$ in the middle, and $\sim 9$ hours for $\otoc^{(2)}$ with $B$ at the edge, for computing all observable locations of a single random circuit instance on a single NVIDIA H200 GPU. We expect significant improvements of the runtime with further optimizations, especially for $\otoc^{(2)}$, by using techniques such as grouping tensor network contractions with shared tensors.

The error and SNR benchmark of our algorithm is computed by comparing against exact $\otoc$ values, which are obtained by exact tensor network contraction on GPUs using complex64 precision. These exact contractions have exponential cost in general. For the data presented in Fig.~\ref{fig:otoc_numerics}, a single exact observable value takes up to $\sim2$ GPU hours. At this scale, certain exact observables are already hard to compute. For example, Fig.~\ref{fig:otoc_numerics} shows $\otoc^{(1)}$ benchmarks at location $x=27$, which is the leftmost location that we can compute exactly. At location $x=25$, we are not able to find a contraction path that makes the exact computation feasible, while we expect our algorithm to still give high-accuracy approximations. This shows that our algorithm gives practically efficient and accurate $\otoc$ observables that can be much harder to compute using other methods.

\section{Outlook}
\label{sec:outlook}

Our work provides a new class of classical algorithms to efficiently compute observables involving multiple copies of forward and backward chaotic time evolution. These  algorithms exploit the physical phenomenon of chaotic dephasing, which leads to two useful features: (i) there exists a simple ``permutation'' basis, where we can organize such observables into an easily computable leading contribution and a hierarchy of subleading contributions, and (ii) the subleading contributions are increasingly suppressed when they involve more computationally intensive longer time intervals.  Efficient computation is possible if the discarded terms decay sufficiently rapidly, and if the retained contributions can be efficiently resummed. Thus,  a true quantum advantage task must evade such an ``easily summable basis''.

OTOCs provide a prototype application for our algorithm, but the membrane approach has much broader applicability~\cite{nahumQuantumEntanglementGrowth2017,zhouEmergentStatisticalMechanics2019,zhou_entanglement_2020,mezeiMembraneTheoryEntanglement2018,liEntanglementDomainWalls2023,gongCoarseGrainedEntanglementOperator2022,sommersZeroTemperatureEntanglementMembranes2024,sierantEntanglementGrowthMinimal2023,haRougheningTransitionQuantum2025}. For instance, the method is applicable to other multi-copy observables involving forward and backward time evolution which often occur in quantum information protocols, such as subsystem purity. Moreover, the circuits need not be random: since we only rely on an assumption about chaotic dephasing, we expect the algorithm to be effective for locally scrambling circuits with translation invariance in space and time. 

While our approach is stated as an algorithm for classical simulation, the ideas we have introduced can also be adapted to hybrid quantum-classical simulation. Access to inputs from smaller-scale quantum devices could further boost the practical efficiency of our algorithm: using a hybrid approach, one can compute the ``bubble'' contributions on a quantum device, and plug them into our algorithm to obtain $\otoc^{(k)}$ values for much larger system sizes and times.

Even when the algorithm does not achieve polynomial scaling, it provides a way to extend finite-size calculations to larger many-body systems, much like modern tensor network methods. Thus, even without favorable scaling in two spatial dimensions, our algorithm offers a practical route to benchmarking two-dimensional ${\otoc}^{(2)}$ experiments.

For physical applications, this is the first practical algorithm to compute not only ensemble averages but also circuit-to-circuit fluctuations and higher-order moments of observables in random quantum circuits. Higher order statistics of observables~\cite{mccullochFullCountingStatistics2023} have been used to detect asymmetry in the tails of the distribution~\cite{Rosenberg2024}, which can distinguish different universality classes sharing the same two-point correlations~\cite{gopalakrishnanDistinctUniversalityClasses2024,krajnikUniversalAnomalousFluctuations2024,takeuchiPartialYetDefinite2025}. Our algorithm provides a practical way to perform ``experiments'' on these systems via classical simulation.

The precise scope of our algorithm remains to be fully determined. In particular, it is an open question whether it can be modified to efficiently compute $\otoc$s on circuits on ${\rm 2D}$ geometries.
Higher dimensions make the resummation more difficult, even though we expect that long-duration contributions remain rapidly suppressed by chaotic dephasing. 
Finally, it remains to understand how conservation laws affect the applicability of the algorithm to Hamiltonian evolution, and whether a modified basis may facilitate classical computation in that setting~\cite{khemaniOperatorSpreadingEmergence2018,rakovszkyDiffusiveHydrodynamicsCorrelators2018,rakovszkySubBallisticGrowthRenyi2019,jonaySlowThermalizationSubdiffusion2024,chenQuantumButterflyEffect2020,marvianRestrictionsRealizableUnitary2022}.

{\bf Acknowledgements.} We thank KITP for its hospitality during the program ``Learning the Fine Structure of Quantum Dynamics in Programmable Quantum Matter'' in October 2025, when this project began. This research was supported in part by grant NSF PHY-2309135 to the Kavli Institute for Theoretical Physics (KITP).  SV acknowledges funding provided by the DOE QuantISED program (DE SC0018407) and the Institute for Quantum Information and Matter, an NSF Physics Frontiers Center (NSF Grant PHY- 2317110). YL acknowledges compute resources provided by the Blue Vela cluster of IBM Research. TZ acknowledges Adam Nahum for helpful discussions and previous collaboration in closely related projects. TZ acknowledges Advanced Research Computing (ARC) at Virginia Tech for providing the cluster resources used in this work.

{\bf AI disclosure.} All key ideas, and a complete theoretical formulation of the membrane cluster expansion algorithm and its guarantees, were developed by hand. We then used OpenAI's Codex to assist with coding and implementation of our algorithm. 

\appendix 
\section{Additional details on the membrane cluster expansion  algorithm}\label{app:algo}

\subsection{ Permutation states and boundary states for OTOCs}
\label{app:otoc1_definitions}
\label{app:higher_point}

Given some operator $O$ acting on one copy of a Hilbert space $\sH$ of one qubit, we can define the following set of states on $2m$ copies $(\sH \otimes \bar\sH)^{\otimes m}$, associated with $O$ and any permutation $\sigma \in \sS_m$:   
\begin{equation}
\braket{i_1i_1'... i_m i_m'|O, \sigma} = O_{i_mi'_{\sigma(m)}} ... O_{i_mi'_{\sigma(m)}}    
\end{equation}
where $\{\ket{i_a}\}$ is a basis for the $a$'th copy of $\sH$, and $\{\ket{i'_a}\}$ for the $a$'-th copy of $\bar H$.
We label the special set of states for the case where $O$ is the identity operator as $\ket{\sigma}$. 

For any $k$, $\otoc^{(k)}$ can be expressed as a transition amplitude in $(\sH \otimes \bar \sH)^{\otimes 2k}$ as stated in \eqref{otoc_1_amp}, with 
\begin{align}
&|\Psi_{\rm in}(B_0)\rangle = \ket{B, e}_0 \ket{e}_1 ... \ket{e}_{n-1} \\
&\out = \ket{\eta}_0 ... \ket{\eta}_{x-1}\ket{M, \eta}_{x} \ket{\eta}_{x+1}... \ket{\eta}_{n-1}
\end{align}
where the subscripts indicate ($4k$ copies of) lattice sites, and $\eta$ is the cyclic permutation on $2k$ elements given by $\eta= (1 \, 2\, ...  \, 2k)$.

For the $k=1$ case, we have only two permutation states corresponding to $\sigma \in \sS_2$, and we have given them the special labels $\ket{+}= \ket{e}$ and $\ket{-} = \ket{(12)}$.

\subsection{Dual states for $\otoc^{(1)}$} 

\label{app:duals}

Let us now explicitly define the dual states for the states in the simple subspace for $\otoc^{(1)}$, which enter into the definitions of the projector $\hat d_t$, as well as the amplitudes $W$ and $Z$.

Recall that in the geometry of Fig.~\ref{fig:circuit_setup}, odd layers contain only
2-site blocks. On even layers, the leftmost block has 3 sites and all
others have 2 sites (see \eqref{blocks} below for an example at $t=10$). 

Let us start with the $k=1$ case. On a gate block of $p$ qubits, the pairing states $\ket{+}^{\otimes p}$ and
$\ket{-}^{\otimes p}$ each have squared norm $2^{2p}$, and their mutual overlap is $2^p$.
Inverting their Gram matrix gives the block dual
\begin{equation}
\ket{(\sigma^{\otimes p})^*}
=\frac{\ket{\sigma}^{\otimes p}-2^{-p}\ket{\bar\sigma}^{\otimes p}}{2^{2p}-1},
\qquad \sigma\in\{+,-\},
\end{equation}
where $\bar\sigma$ is the opposite pairing. These dual states are  defined such that 
\begin{equation}
    \braket{\sigma^{\otimes p}| ({\sigma'}^{\otimes p})^{\ast}} = \delta_{\sigma \sigma'} \, .
\end{equation} 

Hence, the projectors onto given permutation states for one block are given by 
\begin{equation}
    P_{\sigma\sigma } = \ket{(\sigma \sigma )^{\ast}}\bra{\sigma \sigma }, \quad P_{\sigma\sigma \sigma} = \ket{(\sigma \sigma \sigma)^{\ast}}\bra{\sigma \sigma \sigma}
\end{equation} 
depending on whether the block consists of 2 or 3 sites. The dual of the single domain 
wall state $\ket{D_x}$ of Sec.~\ref{sec:db} is then the  product of minus block
duals to the left of $x$, and plus block duals to its right. The two extreme
wall positions give the all-plus and all-minus states. Blockwise duality then
gives $\langle D_x|D_y^\ast\rangle=\delta_{xy}$.

When a term $\ket{D_x^{\ast}}\bra{D_x}$  in the projector $\hat d_t$  is inserted below the corresponding unitary layer $U_t$, it is left invariant. This follows immediately from the above definitions, and the unitarity of each gate. We show an example below for $t=10$, and $x$ between sites 4 and 5:  
\begin{equation}
\begin{aligned}
\hat d_x &=
\begin{tikzpicture}[scale=0.8,baseline=-0.25cm]
  \begin{scope}[line width=0.7pt,line cap=butt,line join=miter]
    \def\wiresep{0.8}
    \def\gatepad{0.2}
    \foreach \s/\n/\c/\lab in {
      0/3/red/{---},
      3/2/red/{--},
      5/2/blue/{++},
      7/2/blue/{++},
      9/2/blue/{++}
    }{
      \pgfmathsetmacro{\blockshift}{\s*\wiresep}
      \pgfmathsetmacro{\gatewidth}{(\n-1)*\wiresep+2*\gatepad}
      \begin{scope}[xshift=\blockshift cm]
        \foreach \j in {1,...,\n}{
          \pgfmathsetmacro{\wirex}{\gatepad+(\j-1)*\wiresep}
          \draw (\wirex,0.45) -- (\wirex,0.05);
          \draw (\wirex,-0.55) -- (\wirex,-1.26);
        }
        \draw (0,-0.55) rectangle (\gatewidth,0.05);
        \path[draw=black,fill=\c]
          (0.1,-1.1) rectangle ({\gatewidth-0.1},-0.85);
        \node[anchor=north,inner sep=1pt]
          at ({\gatewidth/2},-1.4) {$P_{\lab}$};
      \end{scope}
    }
    \pgfmathsetmacro{\cutx}{\gatepad+4.5*\wiresep}
    \draw[dashed] (\cutx,-1.3) -- (\cutx,0.65)
      node[above] {$x$};
  \end{scope}
\end{tikzpicture}
\\[1.5ex]
&=
\begin{tikzpicture}[scale=0.8,baseline=-0.05cm]
  \begin{scope}[line width=0.7pt,line cap=butt,line join=miter]
    \def\wiresep{0.8}
    \def\gatepad{0.2}
    \path (0,0) ({10*\wiresep+2*\gatepad},0);
    \foreach \s/\n/\c/\lab in {
      0/3/red/{---},
      3/2/red/{--},
      5/2/blue/{++},
      7/2/blue/{++},
      9/2/blue/{++}
    }{
      \pgfmathsetmacro{\blockshift}{\s*\wiresep}
      \pgfmathsetmacro{\gatewidth}{(\n-1)*\wiresep+2*\gatepad}
      \begin{scope}[xshift=\blockshift cm]
        \foreach \j in {1,...,\n}{
          \pgfmathsetmacro{\wirex}{\gatepad+(\j-1)*\wiresep}
          \draw (\wirex,0.45) -- (\wirex,-0.285);
        }
        \path[draw=black,fill=\c]
          (0.1,-0.125) rectangle ({\gatewidth-0.1},0.125);
        \node[anchor=north,inner sep=1pt]
          at ({\gatewidth/2},-0.425) {$P_{\lab}$};
      \end{scope}
    }
    \pgfmathsetmacro{\cutx}{\gatepad+4.5*\wiresep}
    \draw[dashed] (\cutx,-0.325) -- (\cutx,0.65)
      node[above] {$x$};
  \end{scope}
\end{tikzpicture}
\end{aligned}
\label{blocks}
\end{equation}

\subsection{Two domain wall subspace for operator in the middle}
\label{app:two_dw}

When $B_0$ lies in the interior of the chain, its light cone grows on both sides.
Layer $t$ of the pyramid, counted from the top, contains $t$ two-qubit gates.
On the cut immediately below that layer, we group the $2t$ active sites into
the $t$ pairs acted on by these gates. We label the boundaries of the pairs
by $0,\ldots,t$ from left to right. The simple subspace on this cut is
spanned by the states $\ket{D_{x_1,x_2}}$ with $0\le x_1<x_2\le t$, defined by
\begin{equation}
\ket{D_{x_1,x_2}}
= \ket{++}^{\otimes x_1}\,
  \ket{--}^{\otimes (x_2-x_1)}\,
  \ket{++}^{\otimes (t-x_2)} .
\label{eq:two_dw_interval}
\end{equation}
Here $\ket{++}=\ket{+}\otimes\ket{+}$ and likewise for $\ket{--}$.
The walls at $x_1$ and $x_2$ specify a single nonempty $\ket{-}$ domain. This domain may reach either edge of the system or fill it entirely. Coincident walls would give the all-$\ket{+}$ configuration, which is excluded from the simple subspace as it is expected to give an exponentially small contribution to the OTOC. Thus $\dim\mathcal S_t=\binom{t+1}{2}$.

To obtain the dual states $\ket{(D_{x_1,x_2})^*}$, replace each block in Eq.~\eqref{eq:two_dw_interval} by its two-site dual ($p=2$) from Appendix~\ref{app:otoc1_definitions}. The resulting product states
satisfy $\langle D_{x_1,x_2}|(D_{y_1,y_2})^*\rangle
=\delta_{x_1y_1}\delta_{x_2y_2}$, so the insertion at layer $t$ is
\begin{equation}
\hd_t=\sum_{0\le x_1<x_2\le t}
\ket{(D_{x_1,x_2})^*}\bra{D_{x_1,x_2}} .
\end{equation}
With $\vec s=(x_1,x_2)$, the expansion of Sec.~\ref{sec:otoc1} applies directly. Each bulk weight $W(t,\vec s;t',\vec s\,')$ connects two interval states over a common duration $t'-t$ for both walls. The physical operators $B_0$ and $M_x$ supply the boundary conditions at the top and bottom of the
pyramid, respectively. The algorithm for computing $\otoc^{(1)}_\Delta(T,x)$ proceeds as in the
single domain wall case of Sec.~\ref{sec:otoc1}, with the sum over intermediate states running over the allowed wall pairs $(x_1,x_2)$ at each insertion.

\subsection{The simple subspace and dual states for $\otoc^{(2)}$}
\label{app:otoc2}

Recall that in our circuit geometry, on time slice $t$, below layer $t$ of the flat-edge circuit, the $t+1$ sites form $m=\lceil t/2\rceil$ gate blocks. 
Odd layers contain only two-site blocks, while on even layers the leftmost block has three sites and all others have two. Let $p_j$ be the size of block $j$, numbered from left to right. We first repeat the definition of the simple states in Sec.~\ref{sec:otoc2} more formally.  A simple state  is a subset of states with one permutation $\rho_j\in\sS_4$ per block:
\begin{equation}
\ket{\vec s}=\bigotimes_{j=1}^m\ket{\rho_j}^{\otimes p_j},
\qquad \vec s=(\rho_1,\ldots,\rho_m).
\end{equation}
To specify the states in the simple subspace, set $\rho_0=\eta=(1234)$ and $\rho_{m+1}=e$, the identity permutation. The Cayley distance $d(\pi,\rho)=4-c(\pi^{-1}\rho)$ is the minimum number of transpositions between the two permutations, where $c$ counts the number of cycles in the permutation, including fixed points. We include precisely the permutation profiles $\vec s$ that follow a shortest path from $\eta$ to $e$, that is,
\begin{equation}
\sum_{j=0}^{m}d(\rho_j,\rho_{j+1})=d(\eta,e)=3.
\label{eq:otoc2_geodesic_profiles}
\end{equation}
In the wall notation of Sec.~\ref{sec:otoc2}, the positions $x_3\le x_2\le x_1$ lie at block boundaries, including the two ends of the cut. A jump of distance two or three is allowed, and represents coincident ``elementary domain walls''. The leftmost $\eta$ domain and the rightmost $e$ domain may each be empty. For example, the uniform $e$ profile has all three walls at the left edge. Each distinct block profile is counted once, including when a jump admits several factorizations into transpositions. The number of profiles with one, two, or three distinct jumps is, respectively, $m+1$, $12\binom{m+1}{2}$, and $16\binom{m+1}{3}$, so the dimension of the simple subspace grows as $O(t^3)$.

As in Appendix~\ref{app:otoc1_definitions}, we obtain the dual of a simple state $\ket{\vec s}$ by taking the product of its block duals. For $\otoc^{(2)}$, the Gram matrix on a block of $p$ sites has entries $G^{[p]}_{\pi\rho}=2^{p\,c(\pi^{-1}\rho)}$. The dual of $\ket{\rho}^{\otimes p}$ is
\begin{equation}
\ket{(\rho^{\otimes p})^*}
=\sum_{\pi\in\sS_4}\big[(G^{[p]})^{-1}\big]_{\pi\rho}\ket{\pi}^{\otimes p},
\end{equation}
where the sum runs over all 24 permutations. The insertion $\hd_t$ is then the sum of $\ket{\vec s^{\,*}}\bra{\vec s}$ over the profiles satisfying Eq.~\eqref{eq:otoc2_geodesic_profiles}.

\subsection{Top and bottom boundary treatment}
\label{app:boundary_layer}
\label{app:top}

\begin{figure}[t]
\centering
\includegraphics[width=\columnwidth]{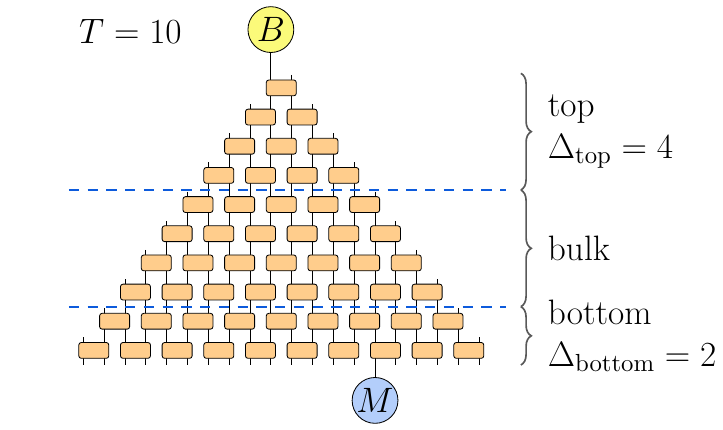}
\caption{Top and bottom boundary regions in a pyramid circuit of depth $T=10$. The dashed lines lie below the fourth row from the top and above the second row from the bottom, corresponding to $\Delta_{\mathrm{top}}=4$ and $\Delta_{\mathrm{bottom}}=2$. The top insertion can lie on any of layers $1$ through $4$, and the bottom insertion on layer $8$ or $9$. Boundary amplitudes are evaluated exactly, while the intervening region is treated using the bulk cluster expansion.}
\label{fig:boundary_regions}
\end{figure}

In this section, we describe the top and bottom boundary treatment of the membrane cluster expansion algorithm. We use the pyramid geometry as an example. Fig.~\ref{fig:boundary_regions} shows a circuit of depth $T=10$, with the top and bottom boundary regions separated from the bulk by dashed lines. The bulk algorithm proceeds as described in Sec.~\ref{sec:algorithm}.

We label the layers $1,2,\ldots,T$ from top to bottom, so the replicated evolution is $\mathcal U_1\cdots\mathcal U_T$, with $B_0$ at the top and $M_x$ at the bottom. We expand $\mathcal U_t=\hd_t+\hb_t$ on layers $1,\ldots,T-1$ and leave $\mathcal U_T$ unexpanded. For a simple state $\ket{\vec s_t}$ at layer $t$, the corresponding term in $\hd_t$ is $\ket{\vec s_t^{\,*}}\bra{\vec s_t}$, where $\ket{\vec s_t^{\,*}}$ is its dual. For a history with a $\hd$ insertion in each boundary window, let $j$ be its last insertion in the top window and $\ell$ its first in the bottom window. We evaluate the boundary amplitudes
\begin{align}
Z_{\mathrm{top}}(j,\vec s_j)
&=\inst\mathcal U_1\cdots\mathcal U_{j-1}\ket{\vec s_j^{\,*}},
\label{eq:boundary_Z_top}\\
Z_{\mathrm{bottom}}(\ell,\vec s_\ell)
&=\bra{\vec s_\ell}\mathcal U_{\ell+1}\cdots\mathcal U_T\out,
\label{eq:boundary_Z_bottom}
\end{align}
by exact tensor network contraction. An empty product of evolution operators is the identity. Thus for fixed $\vec s_j$ and $\vec s_\ell$, $Z_{\mathrm{top}}$ sums all histories above $j$, and $Z_{\mathrm{bottom}}$ sums all histories below $\ell$.

In the exact expansion, successive $\hd$ insertions at layers $t_1<t_2$ are connected by the irreducible amplitude
\begin{equation}
W(t_1,\vec s_1;t_2,\vec s_2)
=\bra{\vec s_1}\hb_{t_1+1}\cdots\hb_{t_2-1}\ket{\vec s_2^{\,*}}.
\label{eq:boundary_W}
\end{equation}
Choosing $j$ and $\ell$ in this way counts each history that reaches both windows once. For disjoint windows, $\Delta_{\mathrm{top}}+\Delta_{\mathrm{bottom}}<T$, the explicit insertions lie at layers $j=t_1<\cdots<t_m=\ell$, with
\begin{equation}
\begin{aligned}
1&\le t_1\le\Delta_{\mathrm{top}},\\
\Delta_{\mathrm{top}}&<t_i<T-\Delta_{\mathrm{bottom}}\quad(1<i<m),\\
T-\Delta_{\mathrm{bottom}}&\le t_m\le T-1.
\end{aligned}
\label{eq:boundary_windows}
\end{equation}
Let $\mathcal C_m$ denote all such sequences of $m\ge2$ insertions, including the choice of simple state $\vec s_i$ at each layer $t_i$. The regrouped expansion is
\begin{equation}
\begin{aligned}
\otoc^{(1)}(T,x)
&=R_{\mathrm{bd}}+2^{-n}\sum_{m\ge2}\sum_{\mathcal C_m}
Z_{\mathrm{top}}(t_1,\vec s_1)\\
&\quad\times\left[\prod_{i=1}^{m-1}W(t_i,\vec s_i;t_{i+1},\vec s_{i+1})\right]\\
&\quad\times Z_{\mathrm{bottom}}(t_m,\vec s_m).
\end{aligned}
\label{eq:boundary_resummed_expansion}
\end{equation}
Here $R_{\mathrm{bd}}$ is the residual boundary contribution from histories with no $\hd$ insertion in at least one of the two windows. To obtain $\otoc^{(1)}_\Delta(T,x)$, we omit $R_{\mathrm{bd}}$ and replace each $W$ by $W_\Delta$ of Eq.~\eqref{eq:W_truncated}. This includes precisely the sequences with $t_{i+1}-t_i\le\Delta$ for every $1\le i<m$, while summing the histories above $j$ and below $\ell$ exactly.

In practice, we choose $\Delta_{\mathrm{bottom}}=\Delta$ and $\Delta_{\mathrm{top}}\ge\Delta$. A larger top region is useful for two reasons. First, in the flat edge geometry, bubbles attached to the left spatial boundary cause larger truncation errors, and a larger top region includes more of these contributions in the exact boundary contraction. Second, evaluating $Z_{\mathrm{top}}$ costs fewer resources than the corresponding bottom contraction, especially in the flat edge geometry, making a larger top cutoff more affordable computationally.

\section{Error analysis for the membrane cluster expansion algorithm}
\subsection{Sufficient conditions for inverse polynomial relative error}
\label{app:total_error_bound}
Fix total time $T$ and a circuit instance. We now explicitly state two sufficient conditions that guarantee that our algorithm's estimates of $Z(\mathbf{i}; \mathbf{j})$ between any two points in the bulk are accurate up to $1/\operatorname{poly}(T)$ relative error. The conditions we identify in this section apply to estimation of the bulk part of $\otoc^{(k)}$ for arbitrary $k$. In subsequent sections, we give theoretical and numerical justifications that these assumptions are satisfied.

\begin{assumption}[Bubble suppression]\label{assm:1}
There exists a constant $\lambda>0$ and a polynomially bounded
function $A(T)$ such that, for all $1\leq t_1<t_2\leq T$ and
$x_1\in\mathcal X_{t_1}$, $x_2\in\mathcal X_{t_2}$,
\begin{equation}
\left|W(t_1,x_1;t_2,x_2)\right|
\leq A(T)e^{-\lambda(t_2-t_1)}
\left|Z(t_1,x_1;t_2,x_2)\right|.
\end{equation}
\end{assumption}

\begin{assumption}[Convexity-type bound]\label{assm:2}
There exists a polynomially bounded function $C(T)$ such that,
for any $t_1\leq s\leq u\leq t_2$ and configurations
$x_1\in\mathcal X_{t_1}$, $y\in\mathcal X_s$,
$z\in\mathcal X_u$, and $x_2\in\mathcal X_{t_2}$,
\begin{equation}
\begin{aligned}
&|Z(t_1,x_1;s,y)|
 |Z(s,y;u,z)|
 |Z(u,z;t_2,x_2)|\\
&\qquad\leq
C(T)|Z(t_1,x_1;t_2,x_2)|.
\end{aligned}
\end{equation}
\end{assumption}

\begin{theorem}[Relative error guarantee]\label{thm:relerr}
Fix a circuit realization for which \Cref{assm:1,assm:2}
hold. Suppose furthermore that the number of configurations
in the simple subspace $\mathcal{S}_t$ at each time slice is polynomially
bounded in $T$, i.e.,
\begin{equation}
m_T:=\max_{1\leq t\leq T}|\mathcal{S}_t|
=\operatorname{poly}(T).
\end{equation}
Then, in the membrane cluster expansion algorithm for estimating $\otoc^{(k)}$ for a fixed order $k$, there
exists a cutoff
\begin{equation}
\Delta=O\left(\log T+\log\frac{1}{\epsilon}\right)
\end{equation}
such that truncating all bubbles of duration greater than
$\Delta$ yields a relative error of at most $\epsilon$
for all bulk propagators with nonzero exact amplitude:
\begin{equation}
\max_{\substack{
1\leq t_1<t_2\leq T\\
x_1\in\mathcal X_{t_1},\,
x_2\in\mathcal X_{t_2}\\
Z(t_1,x_1;t_2,x_2)\neq 0
}}
\frac{
|Z_\Delta(t_1,x_1;t_2,x_2)
-Z(t_1,x_1;t_2,x_2)|
}{
|Z(t_1,x_1;t_2,x_2)|
}
\leq\epsilon.
\end{equation}
Moreover, whenever $Z(t_1,x_1;t_2,x_2)=0$, the
truncated approximation also vanishes.
\end{theorem}
Note that this theorem holds independently of the efficiency of computing $Z_{\Delta}$. Our choices of $\mathcal{S}_t$ for $\otoc^{(1)}$ and $\otoc^{(2)}$ satisfy the condition that $m_T=\operatorname{poly}(T).$

\begin{proof}[Proof of \Cref{thm:relerr}]

To express the recursion relation \Cref{eq:Z_recursion}
in matrix form, we introduce composite indices
$i=(t_i,\vec s_i)$, where $1\leq t_i\leq T$ and
$\vec s_i\in\mathcal S_{t_i}$ labels a state in the
simple subspace at time $t_i$. We define
\begin{equation}
\mathbf Z_{ij}
:=Z(t_i,\vec s_i;t_j,\vec s_j),
\qquad
\mathbf W_{ij}
:=W(t_i,\vec s_i;t_j,\vec s_j).
\end{equation}
By convention, for any time slice $t$,
\begin{equation}
Z(t,\vec s;t,\vec s\,')
=\delta_{\vec s,\vec s\,'},
\qquad
\vec s,\vec s\,'\in\mathcal S_t,
\end{equation}
and both $Z(t_i,\vec s_i;t_j,\vec s_j)$ and
$W(t_i,\vec s_i;t_j,\vec s_j)$ vanish when
$t_i>t_j$. We also set
\begin{equation}
W(t_i,\vec s_i;t_j,\vec s_j)=0
\qquad\text{if }t_i=t_j.
\end{equation}
Thus, ordering the composite indices by time makes
$\mathbf W$ strictly upper triangular. The recursion \Cref{eq:Z_recursion} is
\begin{equation}\label{eq:recursion}
\mathbf Z_{ij}
=\delta_{ij}
+\sum_k\mathbf W_{ik}\mathbf Z_{kj}.
\end{equation}
In matrix notation, this becomes
\begin{equation}\label{eq:recursion_mat}
\mathbf Z=\id+\mathbf W\mathbf Z.
\end{equation}
Since there are $T$ time slices,
$\mathbf W^{T+1}=0$. Solving
\Cref{eq:recursion_mat} for $\mathbf Z$ gives
\begin{equation}
\mathbf Z
=(\id-\mathbf W)^{-1}
=\id+\mathbf W+\cdots+\mathbf W^T.
\end{equation}

Our algorithm approximates $\mathbf W$ by the
truncated matrix $\mathbf W_\Delta$, defined by
\begin{equation}
(\mathbf W_\Delta)_{ij}
=
\begin{cases}
\mathbf W_{ij},
&1\leq t_j-t_i\leq\Delta,\\
0,&\text{otherwise}.
\end{cases}
\end{equation}
The corresponding approximation $\mathbf Z_\Delta$
satisfies
\begin{equation}
(\mathbf Z_\Delta)_{ij}
=\delta_{ij}
+\sum_k(\mathbf W_\Delta)_{ik}
(\mathbf Z_\Delta)_{kj},
\end{equation}
or, equivalently,
\begin{equation}\label{eq:Zd}
\mathbf Z_\Delta
=\id+\mathbf W_\Delta\mathbf Z_\Delta.
\end{equation}
Hence,
\begin{equation}
\mathbf Z_\Delta
=(\id-\mathbf W_\Delta)^{-1}
=\id+\mathbf W_\Delta+\cdots+
\mathbf W_\Delta^T.
\end{equation}

We now bound the approximation error. Define
\begin{equation}
\mathbf E_W
:=\mathbf W-\mathbf W_\Delta,
\qquad
\mathbf E_Z
:=\mathbf Z-\mathbf Z_\Delta.
\end{equation}
Subtracting \Cref{eq:Zd} from
\Cref{eq:recursion_mat}, we obtain
\begin{align}
\mathbf E_Z
&=\mathbf W\mathbf Z
-\mathbf W_\Delta\mathbf Z_\Delta\\
&=\mathbf E_W\mathbf Z
+\mathbf W_\Delta\mathbf E_Z.
\end{align}
Solving for $\mathbf E_Z$ gives
\begin{equation}
\mathbf E_Z
=(\id-\mathbf W_\Delta)^{-1}
\mathbf E_W\mathbf Z
=\mathbf Z_\Delta\mathbf E_W\mathbf Z.
\end{equation}

Writing this identity in terms of individual matrix
entries,
\begin{align}
(\mathbf E_Z)_{ij}
&=\sum_{k,l}
(\mathbf Z_\Delta)_{ik}
(\mathbf E_W)_{kl}
\mathbf Z_{lj}\nonumber\\
&=\sum_{\substack{k,l:\\
t_l-t_k>\Delta}}
(\mathbf Z_\Delta)_{ik}
\mathbf W_{kl}\mathbf Z_{lj}.
\label{eq:error_entries}
\end{align}

Before bounding the relative error, we observe that
$\mathbf Z_\Delta$ vanishes wherever $\mathbf Z$
vanishes. To see this, consider indices $i,j$ with
$t_i<t_j$ and $\mathbf Z_{ij}=0$.
By \Cref{assm:1}, we have $\mathbf W_{ij}=0$.
Moreover, \Cref{assm:2}, applied with the two
intermediate indices equal to $k$, implies
\begin{equation}
|\mathbf Z_{ik}||\mathbf Z_{kj}|
\leq C(T)|\mathbf Z_{ij}|=0.
\end{equation}
Thus, whenever $\mathbf W_{ik}\neq 0$, we must
have $\mathbf Z_{kj}=0$. Induction on the time
separation $t_j-t_i$, using the recursion for
$\mathbf Z_\Delta$, then shows that
$(\mathbf Z_\Delta)_{ij}=0$.

We may therefore define the maximum relative error
over entries with nonzero exact amplitude:
\begin{equation}
e_{\max}
:=\max_{\substack{i,j:\\\mathbf Z_{ij}\neq0}}
\frac{|(\mathbf E_Z)_{ij}|}{|\mathbf Z_{ij}|}.
\end{equation}
The preceding observation implies that, for all $i,j$,
including entries where $\mathbf Z_{ij}=0$,
\begin{equation}
|(\mathbf Z_\Delta)_{ij}|
\leq(1+e_{\max})|\mathbf Z_{ij}|.
\end{equation}

Define
\begin{equation}\label{eq:beta_delta}
\beta_\Delta
:=\max_{\substack{i,j:\\\mathbf Z_{ij}\neq0}}
\sum_{\substack{k,l:\\t_l-t_k>\Delta}}
\frac{
|\mathbf Z_{ik}|
|\mathbf W_{kl}|
|\mathbf Z_{lj}|
}{
|\mathbf Z_{ij}|
}.
\end{equation}
Using \Cref{eq:error_entries} and taking absolute
values, we obtain
\begin{align}
\frac{|(\mathbf E_Z)_{ij}|}{|\mathbf Z_{ij}|}
&\leq
(1+e_{\max})
\sum_{\substack{k,l:\\t_l-t_k>\Delta}}
\frac{
|\mathbf Z_{ik}|
|\mathbf W_{kl}|
|\mathbf Z_{lj}|
}{
|\mathbf Z_{ij}|
}\nonumber\\
&\leq(1+e_{\max})\beta_\Delta.
\end{align}
Taking the maximum over $i,j$ yields
\begin{equation}
e_{\max}\leq(1+e_{\max})\beta_\Delta.
\end{equation}
Whenever $\beta_\Delta<1$, this implies
\begin{equation}\label{eq:emax_beta}
e_{\max}
\leq\frac{\beta_\Delta}{1-\beta_\Delta}.
\end{equation}

It remains to bound $\beta_\Delta$.
By \Cref{assm:1}, for every pair $k,l$ with
$t_l-t_k>\Delta$,
\begin{equation}
|\mathbf W_{kl}|
\leq A(T)e^{-\lambda(t_l-t_k)}
|\mathbf Z_{kl}|.
\end{equation}
Using \Cref{assm:2}, we consequently have
\begin{align}
&\frac{
|\mathbf Z_{ik}|
|\mathbf W_{kl}|
|\mathbf Z_{lj}|
}{
|\mathbf Z_{ij}|
}\nonumber\\
&\quad\leq
A(T)e^{-\lambda(t_l-t_k)}
\frac{
|\mathbf Z_{ik}|
|\mathbf Z_{kl}|
|\mathbf Z_{lj}|
}{
|\mathbf Z_{ij}|
}\nonumber\\
&\quad\leq
A(T)C(T)e^{-\lambda(t_l-t_k)}.
\end{align}
Therefore,
\begin{equation}
\beta_\Delta
\leq A(T)C(T)
\sum_{\substack{k,l:\\t_l-t_k>\Delta}}
e^{-\lambda(t_l-t_k)}.
\end{equation}

Let
\begin{equation}
m_T:=\max_{1\leq t\leq T}|\mathcal S_t|
\end{equation}
denote the maximum number of simple-subspace
configurations on any time slice.
Each time slice contains at most $m_T$
configurations, and there are at most $T^2$
ordered pairs of time slices. Thus,
\begin{align}
\beta_\Delta
&\leq A(T)C(T)m_T^2T^2
e^{-\lambda\Delta}.
\end{align}
Since $A(T)$, $C(T)$, and $m_T$ are all
polynomially bounded in $T$, we conclude that
\begin{equation}\label{eq:beta_bound}
\beta_\Delta
\leq\operatorname{poly}(T)e^{-\lambda\Delta}.
\end{equation}
By \Cref{eq:emax_beta}, it suffices to choose
$\Delta$ such that
\begin{equation}
\beta_\Delta\leq\frac{\epsilon}{1+\epsilon}.
\end{equation}
For constant $\lambda>0$, this is achieved by
\begin{equation}
\Delta
=O\left(\log T+\log\frac{1}{\epsilon}\right)
\end{equation}
to obtain $e_{max}\leq \epsilon.$
\end{proof}

In the remainder of this appendix, we give
theoretical arguments from statistical mechanics
for why bubble suppression should hold in random
circuits on general local architectures, and provide
additional numerical evidence, beyond that presented
in the main text, for its validity in 1D random
circuits.

\subsection{Theoretical arguments for bubble suppression}
\label{app:WZ_decay_argument}

Now, we give a theoretical argument for the exponential decay of $|W/Z|$ in 1D random circuits. We repeat the reasoning in Ref.~\cite{zhou_entanglement_2020} for the large-$q$ limit, where $q$ is the local Hilbert space dimension. We then estimate the probability for it to hold for a vertical domain wall in qubit circuits ($q=2$) using the distribution of $Z$ \cite{zhouEmergentStatisticalMechanics2019} and a numerical bound on the second moment of $W$.

\begin{figure}[t]
\centering
\includegraphics[width=\linewidth]{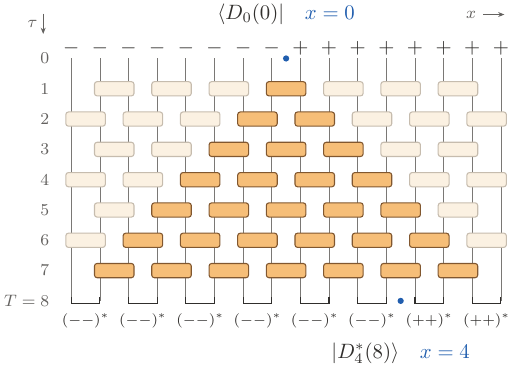}
\caption{Pyramid geometry of $Z(8,4)=Z(0,0;8,4)$. Layers $1,\ldots,7$ are shown from top to bottom. Layer $T=8$ acts trivially on the bottom dual state. Each gate represents its four-copy replicated tensor. Orange gates lie inside the backward light cone of the top wall; pale gates outside cancel by unitarity. The top boundary has $-$ pairings on the left and $+$ pairings on the right, while bottom brackets denote joint two-site dual states forming $|D_4^*(8)\rangle$. Blue dots mark the domain-wall endpoints.}
\label{fig:wz_geometry}
\end{figure}

We consider a 1D brickwork circuit of independent Haar-random two-qudit gates. Using the notations in the main text, the partition function $Z(t,x)$ of interest is
\begin{equation}\label{eq:z_vertical_def}
\begin{aligned}
Z(t,x) &\equiv Z(0,0;t,x) \\
&= \langle D_0(0)|\mathcal U_1\cdots\mathcal U_t|D_x^*(t)\rangle. 
\end{aligned}
\end{equation}
It defines a single domain wall that starts at position $0$ at time $0$ and ends at position $x$ at time $t$. The associated irreducible amplitude $W$ for the same space time displacement is 
\begin{equation}\label{eq:w_vertical_def}
\begin{aligned}
W(t,x) &\equiv W(0,0;t,x) \\
&= \langle D_0(0)|\hb_1\cdots\hb_{t-1}|D_x^*(t)\rangle
\end{aligned}
\end{equation}
for $t> 1$.  For $t = 1$, it is understood that $W(1,x)=Z(1,x)$ . Fig.~\ref{fig:wz_geometry} shows a geometry of the circuit for $t=T=8$ and $x=4$. 

Under Haar averaging, only thin domain-wall paths remain in $Z(t,x)$, with weight $K=q/(q^2+1)$ per step. A path of displacement $x$ has $(t+x)/2$ steps to the right, so the random average is
\begin{equation}
\mathbb{E}[Z(t,x)] = K^t \binom{t}{\frac{t+x}{2}},
\end{equation}
Hence at large $q$, a domain wall of length $t$ carries a leading weight $Z\sim q^{-t}$. 

The irreducible amplitude $W$ starts and ends as a single domain wall but stays outside the single-wall subspace at all intermediate slices. For a segment containing at least one $\hb_\tau$ factor, its Haar average vanishes, $\mathbb E[W]=0$. Due to the conjugation symmetry of the circuit and its boundary state, $W$ is real for each circuit realization. We estimate its magnitude through the second moment $\mathbb E[W^2]$.

Averaging two copies of $W$ gives a permutation spin model taking values in $\mathcal S_4$ and a composite domain wall $(12)(34)$ at both the top and bottom boundaries. The projectors remove any intermediate configuration that contains a single thin wall in either copy. The bulk must therefore contain at least four elementary domain walls at each slice. Each wall costs a leading factor $q^{-1}$ per time step, giving $\mathbb E[W^2]\sim q^{-4t}$ at the leading order in large-$q$. The root-mean-square scale of $W$ is then $q^{-2t}$, suggesting a relative $q^{-t}$ suppression compared with $Z$.

The argument above applies when $q \rightarrow \infty$. For smaller $q$, including $q=2$, we present a more quantitative estimate for a vertical single domain wall, i.e., $x=0$. Our calculation combines a continuum estimate of the distribution of $|Z(t,0)|$ with a numerical bound on $\mathbb E[W^2]$.

Specifically, we estimate the probability that the denominator is unusually small,
\begin{equation}
\delta_Z(t)=\mathbb P\bigl(|Z(t,0)| \lesssim  e^{-\lambda_Zt}\bigr),
\label{eq:delta_Z_def}
\end{equation}
for a rate $\lambda_Z$ larger than the typical decay rate of $Z$. Hereafter,  the symbol $\lesssim$  allows a positive multplicative constant. 

At large $q$, the (bulk of) replicated moments of $Z$ map to those of a directed polymer in a random medium~\cite{zhouEmergentStatisticalMechanics2019}. In the continuum description, the free energy $S_2=- \ln Z$ obeys the Kardar-Parisi-Zhang (KPZ) equation~\cite{nahumQuantumEntanglementGrowth2017},
\begin{equation}
\partial_t S_2=\tilde v_2\ln q+\frac{1}{2}\partial_x^2S_2-\frac{1}{2}(\partial_xS_2)^2+\xi(t,x),
\label{eq:kpz_equation}
\end{equation}
where $\tilde v_2\ln q=\ln[(q^2+1)/(2q)]$. The Gaussian noise has zero mean and covariance
\begin{equation}
\mathbb E[\xi(t,x)\xi(t',x')]\simeq\frac{1}{4q^4}\delta(x-x')\delta(t-t')
\label{eq:polymer_coefficients}
\end{equation}
to leading order at large $q$. The typical late-time growth rate is $v_2\ln q$, with $v_2=\tilde v_2+1/(384q^8\ln q)+\cdots$. Thus the typical and random-average decay rates differ at high order in the large-$q$ expansion.

The scaling of $S_2$ has a crossover time $t_{\mathrm{KPZ}}\sim q^8$. For $1\ll t\ll t_{\mathrm{KPZ}}$, linearizing the fluctuations gives Edwards-Wilkinson (EW) scaling and a Gaussian distribution. At $t\gg t_{\mathrm{KPZ}}$, the nonlinear term leads to KPZ scaling. To leading order in the drift and fluctuation scales,
\begin{equation}
S_2(t)\simeq
\begin{cases}
\tilde v_2\ln q\,t+\mathcal O(q^{-2})t^{1/4}\chi_{\mathrm G}, & 1\ll t\ll t_{\mathrm{KPZ}},\\[6pt]
v_2\ln q\,t-\mathcal O(q^{-8/3})t^{1/3}\chi_{\rm TW}, & t\gg t_{\mathrm{KPZ}},
\end{cases}
\label{eq:typical_s2_cases}
\end{equation}
where $\chi_{\mathrm G}\sim\mathcal N(0,1)$, and the positive coefficients in the $\mathcal O$ terms depend on the boundary conditions. For free boundary condition ($\sum_x Z(t,x)$),  $\chi_{\rm TW}$ follows the Tracy-Widom distribution $F_1$ (GOE); while the fixed boundary condition, $\chi_{\rm TW}$ follows the Tracy-Widom distribution $F_2$ (GUE). The slope $v_2\ln q$ and $\tilde v_2\ln q$ are the growth rates of the typical free energy and of $-\ln\mathbb E[Z]$. 

We choose $\lambda_Z$ larger than both $v_2\ln q$ and $\tilde v_2\ln q$. The event $Z \lesssim e^{-\lambda_Zt}$ is equivalent to $S_2 \gtrsim \lambda_Zt$. Substituting the fluctuation scales in Eq.~\eqref{eq:typical_s2_cases}, this event requires 
\begin{equation}
\begin{cases}
\chi_{\mathrm G}\gtrsim t^{3/4}, & \text{EW},\\
\chi_{\rm TW} \lesssim - t^{2/3}, & \text{KPZ}.
\end{cases}
\label{eq:tail_thresholds}
\end{equation}
The Gaussian upper tail gives a factor $t^{-3/4}\exp[-\mathrm{const.}\,t^{3/2}]$. In the KPZ regime, the left tails $F_1(-u)$ and $F_2(-u)$ have factors $u^{-1/16}$ and $u^{-1/8}$ multiplying $\exp[-\mathrm{const.}\,u^3]$~\cite{baikAsymptoticsTracyWidomDistributions2008}. Substituting $u\sim t^{2/3}$ gives the estimates
\begin{equation}
\begin{aligned}
&\mathbb P\bigl( Z(t) \lesssim e^{-\lambda_Zt}\bigr)\\
&\quad\simeq
\begin{cases}
t^{-3/4}\exp[-\mathrm{const.}\,t^{3/2}], & \text{EW},\\[5pt]
t^{-1/24}\exp[-\mathrm{const.}\,t^2], & \text{KPZ, free endpoint},\\[5pt]
t^{-1/12}\exp[-\mathrm{const.}\,t^2], & \text{KPZ, fixed endpoint},
\end{cases}
\end{aligned}
\label{eq:small_z_cases}
\end{equation}
all giving faster than exponential decay. 

Next we consider $W$. For the following argument, we assume that for all sufficiently large allowed $t$,
\begin{equation}\label{eq:w_second_moment}
\mathbb{E}\bigl[W(t,0)^2\bigr]  \lesssim e^{-2 \lambda_W t}.
\end{equation}
In \Cref{fig:EW_EZ}, we present numerical evidence for this assumption. 

For $\lambda>0$, Markov's inequality applied to $W(t,0)^2$ gives
\begin{equation}\label{eq:markov_bound}
\begin{aligned}
\delta_W(t) = &\mathbb P\bigl(|W(t,0)| \gtrsim e^{-(\lambda_Z+\lambda)t}\bigr)\\
&\lesssim  e^{2(\lambda_Z+\lambda)t}\mathbb E\bigl[W(t,0)^2\bigr]\\
&\lesssim e^{-(2\lambda_W-2\lambda_Z-2\lambda)t}.
\end{aligned}
\end{equation}
The failure probability can decay exponentially if $\lambda_W >  \lambda_Z$. 

If the $Z$-decay has a lower bound $|Z(t,0)|\geq e^{-\lambda_Zt}$ and the $W$-decay has an upper bound $|W(t,0)|<e^{-(\lambda_Z+\lambda)t}$, then $|W(t,0)|<e^{-\lambda t}|Z(t,0)|$. A union bound on the two failure events therefore gives
\begin{equation}\label{eq:union_bound}
\begin{aligned}
&\mathbb{P}\bigl(|W(t,0)| < e^{-\lambda t}|Z(t,0)|\bigr)  \ge 1 - \delta_Z(t) - \delta_W(t).
\end{aligned}
\end{equation}
If the circuit probability $\delta_Z(t)$ decays as rapidly as the polymer estimate in Eq.~\eqref{eq:small_z_cases}, the total failure probability is exponentially small for
\begin{equation}
0 < \lambda <  \lambda_W - \lambda_Z.
\label{eq:lambda_window}
\end{equation}

For qubit system, we can take 
\begin{equation}
\label{eq:lambda_Z}
    \lambda_Z \approx v_2 \ln q \Big|_{q = 2} = \ln \frac{q^2 + 1}{2q}  \Big|_{q = 2} \approx 0.22
\end{equation}
Fig.~\ref{fig:EW_EZ} shows the numerical data of the exponential decays for both $\mathbb{E}[Z^2]$ and $\mathbb{E}[W^2]$. The exponent fit for $\mathbb{E}[W^2]$ suggests taking 
\begin{equation}
    \lambda_W \approx 0.46.
\end{equation}
and the fit for $\mathbb{E}[Z^2]$ is consistent with Eq.~\eqref{eq:lambda_Z}. 

We thus have a window of $\lambda_W -\lambda_Z$ to accommodate a positive $\lambda$.  In summary, the polymer estimate for the lower tail of $|Z|$ and the numerical second moment of $W$ support exponential decay of $|W/Z|$ for a vertical domain wall.

\subsection{Additional numerical evidence for Assumption 1}
\label{app:decay_appendix}
\begin{figure}[htbp]
\centering
\includegraphics[width=0.9\linewidth]{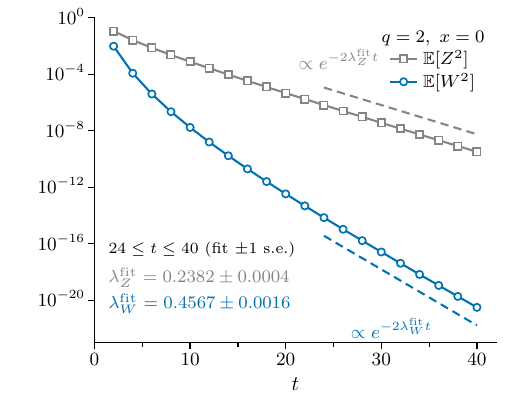}
    \caption{Numerical results of $\mathbb{E}[Z^2]$ and  $\mathbb{E}[W^2]$ for a single domain wall partition function and irreducible weights defined in Eq.~\eqref{eq:z_vertical_def} and Eq.~\eqref{eq:w_vertical_def} . Both show clear exponential decay at late times from $t = 22$ to $t = 40$. Fits in this range give an estimation of $\lambda_W$ and $\lambda_Z$. }
    \label{fig:EW_EZ}
\end{figure}

\Cref{assm:1} is that the following ratio is exponentially suppressed in $|t_l-t_k|$ for the circuit whose $\otoc^{(k)}$ we wish to estimate, for all $k,l$ in the bulk.
\begin{equation}
    \frac{|\mathbf{W}_{kl}|}{|\mathbf{Z}_{kl}|}
\end{equation}
In the main text (Fig.~\ref{fig:w_over_z}) we have plotted the above ratio for our circuit samples, showing that the empirical distribution of these ratios does demonstrate the desired exponential decay with time. As further numerical evidence for \Cref{assm:1} for $\otoc^{(1)}$, we compute the second moments of the numerator and denominator using an adaptation of the variational MPS method \cite{Schollwoeck_2011}, showing in \Cref{fig:EW_EZ} that $\mathbb{E}[W^2]$ and $\mathbb{E}[Z^2]$ both decay exponentially with time. Here, $k$ was chosen to be the spacetime coordinate at the apex of the pyramid (see \ref{fig:wz_geometry}) and $l$ was chosen to be $l$ was chosen to be the spacetime coordinate $(T,0)$. 

\subsection{Additional numerical evidence for Assumption 2}
\begin{figure}
    \centering
    \includegraphics[width=1.0\linewidth]{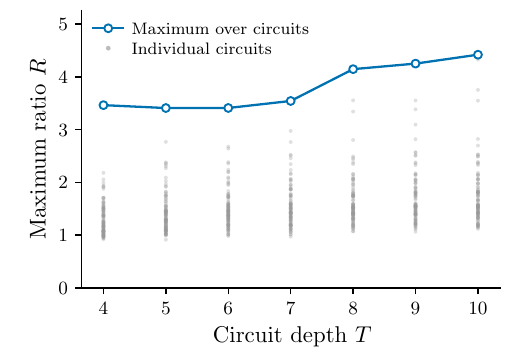}
    \caption{Maxima of the ratio $R$ (see \Cref{eq:R}) in 100 sampled one-dimensional qubit circuits of depths varying from $T=4,\ldots 10$, with Haar-random two-qubit gates. Gray points show each circuit’s maximum value of \(R=|Z_{ik}Z_{kj}|/|Z_{ij}|\) where $i$ was fixed at the circuit apex, $j$ was varied over all spacetime locations at time $T$ and $k$ was varied over all intermediate spacetime locations in the circuit; the blue line connects the maxima over the 100 circuits at each of the time points. $R$ remains of order 1 throughout the entire time interval for these circuits.}
    \label{fig:Assm 2}
\end{figure}

Assumption 2 is that the following ratio grows at most polynomially in $T$ for the circuit whose $\otoc^{(k)}$ we wish to estimate:
\begin{equation}\label{eq:R}
R = \max_{\mathbf{i}, \mathbf{j}, \mathbf{k}} \frac{|Z_{\mathbf{i}\mathbf{k}}||Z_{\mathbf{k},\mathbf{l}}||Z_{\mathbf{l}\mathbf{j}}|}{|Z_{\mathbf{i}\mathbf{j}}|}.
\end{equation}
In \Cref{fig:Assm 2} we present numerical evidence for this assumption for the estimation of $\otoc^{(1)}$. For each of 100 random circuits of total depth $10$, at each intermediate depth \(T=4,\ldots,10\), the initial point \(i\) is fixed at the circuit apex, while \(j\) ranges over all reachable endpoints at time \(t\) and \(k\) over all admissible intermediate times and positions. For each circuit, and for each time point, we compute $R$ and plot it as a gray point in \Cref{fig:Assm 2}. The blue line connects up the maxima over the 100 circuits.

The plot shows that the circuit-specific values of \(R\) remain of order unity throughout the tested depth range, consistent with the assumed polynomial bound on their growth.

\section{Additional data on $\otoc^{(1)}$ and $\otoc^{(2)}$ fluctuations}
\label{app:data_otoc2_fluct}

This appendix presents data of standard deviation computed from the exact values and our approximate algorithm.

\begin{figure*}[!t]
\centering
\includegraphics[width=\textwidth]{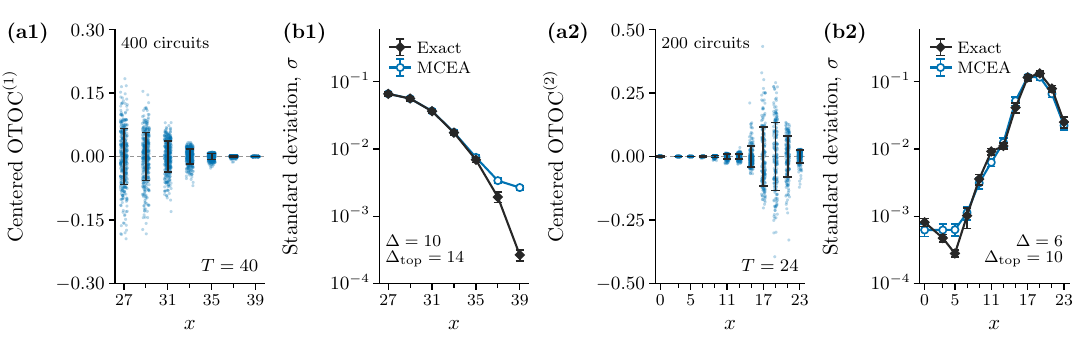}
\caption{\textbf{Spatial dependence of $\otoc^{(1)}$ and $\otoc^{(2)}$ fluctuations.} (a1), (a2) Distributions of the centered $\otoc^{(1)}$ and $\otoc^{(2)}$ for 400 $T=40$ and 200 $T=24$ flat edge circuits, respectively, at all measured positions $x$. Bars show one standard deviation. (b1), (b2) Comparison of the standard deviations computed from the exact circuits with those computed from MCEA with $\Delta=10$ and $\Delta_{\rm top}=14$ for $\otoc^{(1)}$, and $\Delta=6$ and $\Delta_{\rm top}=10$ for $\otoc^{(2)}$. Lines in (b1) and (b2) guide the eye.}
\label{fig:otoc2_fluctuations}
\end{figure*}

For both $\otoc^{(1)}$ and $\otoc^{(2)}$, the circuit-to-circuit fluctuations are appreciable in the sensitive regime $|x|\sim v_B T$ and become small away from it. The agreement between the exact and MCEA standard deviations in Fig.~\ref{fig:otoc2_fluctuations}(b1),(b2) shows that our algorithm provides a practical way to estimate the fluctuation amplitude in the sensitive regime.

The two datasets cover different parts of the fluctuation profile. For $\otoc^{(1)}$ at $T=40$, the available positions $x=27,29,\ldots,39$ lie to the right of $v_B T\simeq24$, and the exact standard deviation decreases towards the edge. MCEA overestimates these small fluctuations at $x=37,39$. For $\otoc^{(2)}$ at $T=24$, the measured positions cover both sides of a peak near $x=19$. MCEA tracks this peak and the decrease towards the edge, while deviations are most visible at small $x$, where the exact fluctuations are weak.

\section{Additional results from implementation of the algorithm}
\label{app:more_results}

\begin{figure*}[!ht]
    \centering
    \includegraphics[width=\linewidth]{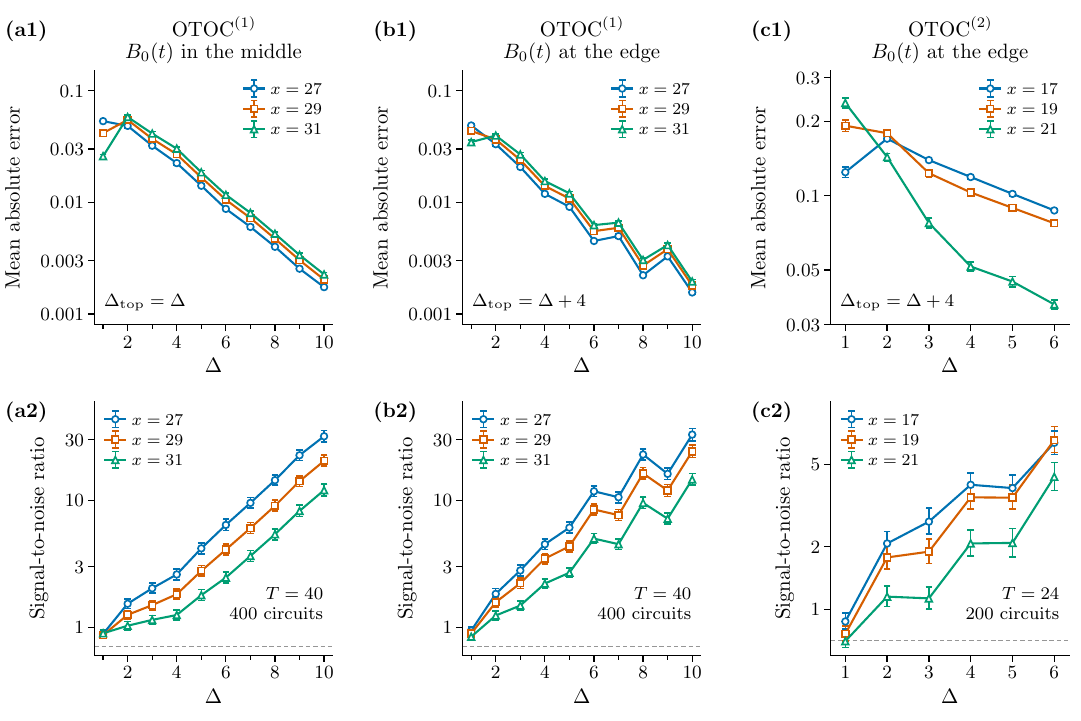} 
    \caption{We show the performance of the membrane cluster expansion algorithm   for various  values of $x$ in the sensitive regime, for both geometries for $\otoc^{(1)}$ \textbf{(a)-(b)}, and for $\otoc^{(2)}$ \textbf{(c)}.}
    \label{fig:sensitive_region}
\end{figure*}

In this Appendix, we show additional data obtained from implementing the membrane cluster expansion algorithm for both $\otoc^{(1)}$ and $\otoc^{(2)}$. In the main text, we highlighted results for a single value of $x$ for each quantity and geometry, lying roughly in the middle of the sensitive regime with significant fluctuations. Here, in Fig.~\ref{fig:sensitive_region}, we show that the decay of the absolute error (top panel) and growth of the SNR (bottom panel) for other $x$ values in the sensitive regime show a similar dependence on $\Delta$. 

\clearpage
\bibliography{biblio}

@article{nahum_op,
  author  = {Nahum, Adam and Vijay, Sagar and Haah, Jeongwan},
  title   = {Operator Spreading in Random Unitary Circuits},
  journal = {Phys. Rev. X},
  volume  = {8},
  number  = {2},
  pages   = {021014},
  year    = {2018},
  doi     = {10.1103/PhysRevX.8.021014},
  eprint  = {1705.08975},
  archivePrefix = {arXiv}
}

@article{Schollwoeck_2011,
   title={The density-matrix renormalization group in the age of matrix product states},
   volume={326},
   ISSN={0003-4916},
   url={http://dx.doi.org/10.1016/j.aop.2010.09.012},
   DOI={10.1016/j.aop.2010.09.012},
   number={1},
   journal={Annals of Physics},
   publisher={Elsevier BV},
   author={Schollwöck, Ulrich},
   year={2011},
   month=Jan, pages={96–192} }

@article{higher_1,
  author = {Fava, Michele and Kurchan, Jorge and Pappalardi, Silvia},
  title = {Designs via Free Probability},
  journal = {Phys. Rev. X},
  volume = {15},
  pages = {011031},
  year = {2025},
  doi = {10.1103/PhysRevX.15.011031},
  eprint = {2308.06200},
  archivePrefix = {arXiv}
}

@article{higher_2,
  author = {Vallini, Elisa and Pappalardi, Silvia},
  title = {Long-time Freeness in the Kicked Top},
  journal = {Quantum},
  volume = {10},
  pages = {2129},
  year = {2026},
  doi = {10.22331/q-2026-06-08-2129},
  eprint = {2411.12050},
  archivePrefix = {arXiv}
}

@article{higher_3,
  author = {Chen, Hyaline Junhe and Kudler-Flam, Jonah},
  title = {Free Independence and the Noncrossing Partition Lattice in Dual-Unitary Quantum Circuits},
  journal = {Phys. Rev. B},
  volume = {111},
  pages = {014311},
  year = {2025},
  doi = {10.1103/PhysRevB.111.014311},
  eprint = {2409.17226},
  archivePrefix = {arXiv}
}

@article{higher_4,
  author = {Jahnke, Viktor and Nandy, Pratik and Pal, Kuntal and Camargo, Hugo A. and Kim, Keun-Young},
  title = {Free Probability Approach to Spectral and Operator Statistics in {Rosenzweig-Porter} Random Matrix Ensembles},
  journal = {JHEP},
  volume = {12},
  pages = {002},
  year = {2025},
  doi = {10.1007/JHEP12(2025)002},
  eprint = {2506.04520},
  archivePrefix = {arXiv}
}

@article{higher_5,
  author = {Fritzsch, Felix and Claeys, Pieter W.},
  title = {Free Probability in a Minimal Quantum Circuit Model},
  journal = {Phys. Rev. X},
  volume = {16},
  pages = {031027},
  year = {2026},
  doi = {10.1103/6mpz-p85s},
  eprint = {2506.11197},
  archivePrefix = {arXiv}
}

@article{higher_6,
  author = {Cotler, Jordan and Hunter-Jones, Nicholas and Liu, Junyu and Yoshida, Beni},
  title = {Chaos, Complexity, and Random Matrices},
  journal = {JHEP},
  volume = {11},
  pages = {048},
  year = {2017},
  doi = {10.1007/JHEP11(2017)048},
  eprint = {1706.05400},
  archivePrefix = {arXiv}
}

@article{higher_7,
  author = {Leone, Lorenzo and Oliviero, Salvatore F. E. and Zhou, You and Hamma, Alioscia},
  title = {Quantum Chaos is Quantum},
  journal = {Quantum},
  volume = {5},
  pages = {453},
  year = {2021},
  doi = {10.22331/q-2021-05-04-453},
  eprint = {2102.08406},
  archivePrefix = {arXiv}
}

@article{higher_8,
  author = {Basu, Pallab and Das, Suman and Nandy, Pratik},
  title = {Complexity of Quadratic Quantum Chaos},
  journal = {JHEP},
  volume = {04},
  pages = {081},
  year = {2026},
  doi = {10.1007/JHEP04(2026)081},
  eprint = {2509.04075},
  archivePrefix = {arXiv}
}

@article{higher_9,
  author = {Haehl, Felix M. and Rozali, Moshe},
  title = {Fine-Grained Chaos in {AdS$_2$} Gravity},
  journal = {Phys. Rev. Lett.},
  volume = {120},
  pages = {121601},
  year = {2018},
  doi = {10.1103/PhysRevLett.120.121601},
  eprint = {1712.04963},
  archivePrefix = {arXiv}
}

@article{higher_10,
  author = {Haehl, Felix M. and Rozali, Moshe},
  title = {Effective Field Theory for Chaotic {CFTs}},
  journal = {JHEP},
  volume = {10},
  pages = {118},
  year = {2018},
  doi = {10.1007/JHEP10(2018)118},
  eprint = {1808.02898},
  archivePrefix = {arXiv}
}

@misc{higher_11,
  author = {Saad, Phil},
  title = {Late Time Correlation Functions, Baby Universes, and {ETH} in {JT} Gravity},
  year = {2019},
  eprint = {1910.10311},
  archivePrefix = {arXiv},
  primaryClass = {hep-th},
  url = {https://arxiv.org/abs/1910.10311}
}

@article{higher_12,
  author = {Yunger Halpern, Nicole and Swingle, Brian and Dressel, Justin},
  title = {Quasiprobability behind the Out-of-Time-Ordered Correlator},
  journal = {Phys. Rev. A},
  volume = {97},
  pages = {042105},
  year = {2018},
  doi = {10.1103/PhysRevA.97.042105},
  eprint = {1704.01971},
  archivePrefix = {arXiv}
}

@article{higher_13,
  author = {Yunger Halpern, Nicole and Bartolotta, Anthony and Pollack, Jason},
  title = {Entropic Uncertainty Relations for Quantum Information Scrambling},
  journal = {Communications Physics},
  volume = {2},
  pages = {92},
  year = {2019},
  doi = {10.1038/s42005-019-0179-8},
  eprint = {1806.04147},
  archivePrefix = {arXiv}
}

@article{Rakovszky2022DAOE,
  author = {Rakovszky, Tibor and von Keyserlingk, C. W. and Pollmann, Frank},
  title = {Dissipation-assisted operator evolution method for capturing hydrodynamic transport},
  journal = {Phys. Rev. B},
  volume = {105},
  pages = {075131},
  year = {2022},
  doi = {10.1103/PhysRevB.105.075131}
}

@misc{Gao2018NoisySimulation,
  author = {Gao, Xun and Duan, Luming},
  title = {Efficient classical simulation of noisy quantum computation},
  year = {2018},
  eprint = {1810.03176},
  archivePrefix = {arXiv},
  primaryClass = {quant-ph},
  url = {https://arxiv.org/abs/1810.03176}
}

@inproceedings{Aharonov2023NoisySampling,
  author = {Aharonov, Dorit and Gao, Xun and Landau, Zeph
            and Liu, Yunchao and Vazirani, Umesh},
  title = {A Polynomial-Time Classical Algorithm for Noisy Random Circuit Sampling},
  booktitle = {Proceedings of the 55th Annual ACM Symposium on Theory of Computing},
  series = {STOC 2023},
  pages = {945--957},
  publisher = {Association for Computing Machinery},
  year = {2023},
  doi = {10.1145/3564246.3585234},
  eprint = {2211.03999},
  archivePrefix = {arXiv}
}

@article{Schuster2025NoisyCircuits,
  author = {Schuster, Thomas and Yin, Chao and Gao, Xun and Yao, Norman Y.},
  title = {A Polynomial-Time Classical Algorithm for Noisy Quantum Circuits},
  journal = {Phys. Rev. X},
  volume = {15},
  pages = {041018},
  year = {2025},
  doi = {10.1103/xct1-7kf2}
}

@article{GonzalezGarcia2025PauliPaths,
  author = {Gonz{\'a}lez-Garc{\'i}a, Guillermo and Cirac, J. Ignacio
            and Trivedi, Rahul},
  title = {{Pauli} path simulations of noisy quantum circuits beyond average case},
  journal = {Quantum},
  volume = {9},
  pages = {1730},
  year = {2025},
  doi = {10.22331/q-2025-05-05-1730},
  eprint = {2407.16068},
  archivePrefix = {arXiv}
}

@article{zhou_entanglement_2020,
  author = {Zhou, Tianci and Nahum, Adam},
  title = {Entanglement Membrane in Chaotic Many-Body Systems},
  journal = {Phys. Rev. X},
  volume = {10},
  pages = {031066},
  year = {2020},
  doi = {10.1103/PhysRevX.10.031066},
  eprint = {1912.12311},
  archivePrefix = {arXiv}
}

@article{vk_op,
  author  = {von Keyserlingk, C. W. and Rakovszky, Tibor
             and Pollmann, Frank and Sondhi, S. L.},
  title   = {Operator Hydrodynamics, {OTOCs}, and Entanglement
             Growth in Systems without Conservation Laws},
  journal = {Phys. Rev. X},
  volume  = {8},
  number  = {2},
  pages   = {021013},
  year    = {2018},
  doi     = {10.1103/PhysRevX.8.021013},
  eprint  = {1705.08910},
  archivePrefix = {arXiv}
}

@article{otoc_rev_1,
  author  = {Larkin, A. I. and Ovchinnikov, Yu. N.},
  title   = {Quasiclassical Method in the Theory of Superconductivity},
  journal = {Sov. Phys. JETP},
  volume  = {28},
  number  = {6},
  pages   = {1200--1205},
  year    = {1969},
  url = {https://jetp.ras.ru/cgi-bin/r/index/e/28/6/p1200?a=list}
}

@article{otoc_rev_2,
  author  = {Almheiri, Ahmed and Marolf, Donald and Polchinski, Joseph
             and Stanford, Douglas and Sully, James},
  title   = {An Apologia for Firewalls},
  journal = {JHEP},
  volume  = {09},
  pages   = {018},
  year    = {2013},
  doi     = {10.1007/JHEP09(2013)018},
  eprint  = {1304.6483},
  archivePrefix = {arXiv},
  primaryClass  = {hep-th}
}

@article{otoc_rev_3,
  author  = {Shenker, Stephen H. and Stanford, Douglas},
  title   = {Black Holes and the Butterfly Effect},
  journal = {JHEP},
  volume  = {03},
  pages   = {067},
  year    = {2014},
  doi     = {10.1007/JHEP03(2014)067},
  eprint  = {1306.0622},
  archivePrefix = {arXiv},
  primaryClass  = {hep-th}
}

@misc{otoc_rev_4,
  author       = {Kitaev, Alexei},
  title        = {Hidden Correlations in the {Hawking} Radiation
                  and Thermal Noise},
  year         = {2014},
  howpublished = {Talk at the Fundamental Physics Prize Symposium},
  note         = {November 10, 2014},
  url          = {https://www.youtube.com/watch?v=OQ9qN8j7EZI}
}

@article{otoc_rev_5,
  author  = {Xu, Shenglong and Swingle, Brian},
  title   = {Scrambling Dynamics and Out-of-Time Ordered Correlators
             in Quantum Many-Body Systems: A Tutorial},
  journal = {PRX Quantum},
  volume  = {5},
  pages   = {010201},
  year    = {2024},
  doi     = {10.1103/PRXQuantum.5.010201},
  eprint  = {2202.07060},
  archivePrefix = {arXiv},
  primaryClass  = {quant-ph}
}

@article{fmi,
  author  = {Vardhan, Shreya and Wang, Jinzhao},
  title   = {Free Mutual Information and Higher-Point
             Out-of-Time-Ordered Correlators},
  journal = {PRX Quantum},
  volume  = {7},
  pages   = {043017},
  year    = {2026},
  doi     = {10.1103/wt7l-6385},
  url     = {https://doi.org/10.1103/wt7l-6385}
}

@Article{Abanin2025,
author={Abanin, Dmitry A.
and Acharya, Rajeev
and Aghababaie-Beni, Laleh
and others},
title={Observation of constructive interference at the edge of quantum ergodicity},
journal={Nature},
year={2025},
month={Oct},
day={01},
volume={646},
number={8086},
pages={825-830},
issn={1476-4687},
doi={10.1038/s41586-025-09526-6},
url={https://doi.org/10.1038/s41586-025-09526-6}
}

@article{mark2024maximum,
  author={Mark, Daniel K. and Surace, Federica and Elben, Andreas and Shaw, Adam L. and Choi, Joonhee and Refael, Gil and Endres, Manuel and Choi, Soonwon},
  title={Maximum entropy principle in deep thermalization and in Hilbert-space ergodicity},
  journal={Physical Review X},
  volume={14},
  number={4},
  pages={041051},
  year={2024},
  publisher={APS},
  doi = {10.1103/PhysRevX.14.041051},
  url = {https://doi.org/10.1103/PhysRevX.14.041051}
}

@article{mcginley2025scrooge,
  author  = {McGinley, Max and Schuster, Thomas},
  title   = {The Scrooge ensemble in many-body quantum systems},
  journal={arXiv preprint arXiv:2511.17172},
  year    = {2025},
  url = {https://arxiv.org/abs/2511.17172}
}

@article{Mok2026Nature,
  title = {Nature Is Stingy: Universality of Scrooge Ensembles in Quantum Many-Body Systems},
  author = {Mok, Wai-Keong and Haug, Tobias and Ho, Wen Wei and Preskill, John},
  journal = {Phys. Rev. X},
  volume = {16},
  issue = {4},
  pages = {041003},
  numpages = {55},
  year = {2026},
  month = {Oct},
  publisher = {American Physical Society},
  doi = {10.1103/tb52-jxmx},
  url = {https://link.aps.org/doi/10.1103/tb52-jxmx}
}

@article{VardhanMoudgalya2026,
  author        = {Vardhan, Shreya and Moudgalya, Sanjay},
  title         = {Entanglement dynamics from universal low-lying modes},
  journal       = {Phys. Rev. B},
  volume        = {113},
  pages         = {014308},
  year          = {2026},
  doi           = {10.1103/prp6-y5hl},
  eprint        = {2407.16763},
  archivePrefix = {arXiv},
  primaryClass  = {cond-mat.stat-mech}
}

@article{jonay,
    title = {Coarse-grained dynamics of operator and state entanglement},
    author = {Jonay, Cheryne and Huse, David A. and Nahum, Adam},
    journal = {arXiv preprint arXiv:1803.00089},
    year = {2018},
    eprint = {1803.00089},
    archivePrefix = {arXiv},
    primaryClass = {cond-mat.stat-mech},
    url = {https://arxiv.org/abs/1803.00089}
}

@article{schuster,
  title = {Learning quantum systems via out-of-time-order correlators},
  author = {Schuster, Thomas and Niu, Murphy and Cotler, Jordan and O'Brien, Thomas and McClean, Jarrod R. and Mohseni, Masoud},
  journal = {Phys. Rev. Research},
  volume = {5},
  pages = {043284},
  year = {2023},
  doi = {10.1103/PhysRevResearch.5.043284},
  publisher = {American Physical Society}
}

@article{cotler,
  title = {Information-theoretic hardness of out-of-time-order correlators},
  author = {Cotler, Jordan and Schuster, Thomas and Mohseni, Masoud},
  journal = {Phys. Rev. A},
  volume = {108},
  pages = {062608},
  year = {2023},
  doi = {10.1103/PhysRevA.108.062608},
  publisher = {American Physical Society}
}

@article{eq_approx,
  author        = {Liu, Hong and Vardhan, Shreya},
  title         = {Entanglement entropies of equilibrated pure states in quantum many-body systems and gravity},
  journal       = {PRX Quantum},
  volume        = {2},
  pages         = {010344},
  year          = {2021},
  doi           = {10.1103/PRXQuantum.2.010344},
  eprint        = {2008.01089},
  archivePrefix = {arXiv}
}

@misc{bermejo2026tensornetworksbeliefpropagation,
      title={Tensor Networks with Belief Propagation Cannot Feasibly Simulate Google's Quantum Echoes Experiment}, 
      author={Pablo Bermejo and Benjamin Villalonga and Brayden Ware and Guifre Vidal and Aaron Szasz},
      year={2026},
      eprint={2604.15427},
      archivePrefix={arXiv},
      primaryClass={quant-ph},
      url={https://arxiv.org/abs/2604.15427}, 
}

@misc{mele2026verifiablequantumadvantagerandom,
      title={Towards verifiable quantum advantage with random circuits: Observables that survive concentration}, 
      author={Antonio A. Mele and Francesco A. Mele and Jarrod R. McClean and Thomas E. O'Brien},
      year={2026},
      eprint={2609.37890},
      archivePrefix={arXiv},
      primaryClass={quant-ph},
      url={https://arxiv.org/abs/2609.37890}, 
}

@Article{Kim2023,
author={Kim, Youngseok
and Eddins, Andrew
and Anand, Sajant
and Wei, Ken Xuan
and van den Berg, Ewout
and Rosenblatt, Sami
and Nayfeh, Hasan
and Wu, Yantao
and Zaletel, Michael
and Temme, Kristan
and Kandala, Abhinav},
title={Evidence for the utility of quantum computing before fault tolerance},
journal={Nature},
year={2023},
month={Jun},
day={01},
volume={618},
number={7965},
pages={500-505},
issn={1476-4687},
doi={10.1038/s41586-023-06096-3},
url={https://doi.org/10.1038/s41586-023-06096-3}
}

@misc{leviatan2026resolvingstructureprethermalfloquet,
      title={Resolving Structure in Prethermal Floquet Dynamics with Precision Quantum Computation}, 
      author={Eyal Leviatan and Tasneem Watad and Roy Perry and Lukas Broers and Mohammed Zuhair Mullath and Ori Alberton and Itai Arad and Yosi Atia and Eyal Bairey and Shaul Barkan and Matan Ben Dov and Asaf Berkovitch and Ewout van den Berg and Itsik Cohen and Omri Golan and Ilya Gurwich and Avieli Haber and Barak A. Katzir and Oded Kenneth and Roei Levi and Yotam Y. Lifshitz and Yaron Lukovsky and Ron Melcer and Adiel Meyer and Boris Muratov and Aviad Panahi and Gili Schul and Tali Shnaider and Maor Shutman and Alireza Seif and Tomonori Shirakawa and Asif Sinay and Vincent P. Su and Hayk Tepanyan and Omri Trebitch and Assaf Zubida and Dorit Aharonov and Hrant Gharibyan and Abhinav Kandala and Seiji Yunoki and Netanel H. Lindner},
      year={2026},
      eprint={2607.24937},
      archivePrefix={arXiv},
      primaryClass={quant-ph},
      url={https://arxiv.org/abs/2607.24937}, 
}

@misc{barron2026observableestimationabsenceclassical,
      title={Observable Estimation in the Absence of Classical Verification}, 
      author={Samantha V. Barron and Bradley Mitchell and Vinay Tripathi and Francesco Grieco and Ilan Rosen and Francesca Pietracaprina and Davide Materia and Alireza Seif and Darvin Wanisch and Ramón L. Panadés-Barrueta and Ewout van den Berg and Jay-U Chung and Andrew Eddins and Sam Ferracin and Guillermo García-Pérez and John Goold and Luke C. G. Govia and Holger Haas and Ian Hincks and Jesse C. Hoke and Zoë Holmes and Su-un Lee and Youngseok Kim and Swarnadeep Majumder and Sabrina Maniscalco and Simone Montangero and Daniel Puzzuoli and Tomaž Prosen and James Raftery and Ricardo Rivera Cardoso and Max Rossmannek and Manuel Rudolph and Brendan Saxberg and Liran Shirizly and Karthik Siva and Joshua Skanes-Norman and Ilaria Siloi and Kevin C Smith and Boris Sokolov and Maika Takita and Yanting Teng and Mao Tian Tan and Joseph Tindall and Zoltán Zimborás and Matteo A. C. Rossi and Minh C. Tran and Sergei N. Filippov and Abhinav Kandala},
      year={2026},
      eprint={2607.25998},
      archivePrefix={arXiv},
      primaryClass={quant-ph},
      url={https://arxiv.org/abs/2607.25998}, 
}

@Article{Arute2019,
author={Arute, Frank
and Arya, Kunal
and Babbush, Ryan
and others},
title={Quantum supremacy using a programmable superconducting processor},
journal={Nature},
year={2019},
month={Oct},
day={01},
volume={574},
number={7779},
pages={505-510},
issn={1476-4687},
doi={10.1038/s41586-019-1666-5},
url={https://doi.org/10.1038/s41586-019-1666-5}
}

@article{Zhong2020,
author = {Han-Sen Zhong  and Hui Wang  and Yu-Hao Deng  and Ming-Cheng Chen  and Li-Chao Peng  and Yi-Han Luo  and Jian Qin  and Dian Wu  and Xing Ding  and Yi Hu  and Peng Hu  and Xiao-Yan Yang  and Wei-Jun Zhang  and Hao Li  and Yuxuan Li  and Xiao Jiang  and Lin Gan  and Guangwen Yang  and Lixing You  and Zhen Wang  and Li Li  and Nai-Le Liu  and Chao-Yang Lu  and Jian-Wei Pan },
title = {Quantum computational advantage using photons},
journal = {Science},
volume = {370},
number = {6523},
pages = {1460-1463},
year = {2020},
doi = {10.1126/science.abe8770},
URL = {https://www.science.org/doi/abs/10.1126/science.abe8770},
}

@misc{roymoulikDQC1hardnessEstimatingCorrelation2024,
  title = {{DQC1}-hardness of estimating correlation functions},
  author = {Roy Moulik, Subhayan and Strelchuk, Sergii},
  year = {2024},
  month = nov,
  eprint = {2411.05208},
  archivePrefix = {arXiv},
  primaryClass = {quant-ph},
  doi = {10.48550/arXiv.2411.05208},
  url = {https://arxiv.org/abs/2411.05208}
}

@misc{kingSimplifiedVersionQuantum2025,
  title = {A simplified version of the quantum {OTOC}$^{(2)}$ problem},
  author = {King, Robbie and Kothari, Robin and Babbush, Ryan and Boixo, Sergio and Kechedzhi, Kostyantyn and O'Brien, Thomas E. and Smelyanskiy, Vadim},
  year = {2025},
  month = oct,
  eprint = {2510.19751},
  archivePrefix = {arXiv},
  primaryClass = {quant-ph},
  doi = {10.48550/arXiv.2510.19751},
  url = {https://arxiv.org/abs/2510.19751}
}

@article{Rosenberg2024,
title	= {Dynamics of magnetization at infinite temperature in a {Heisenberg} spin chain},
author	= {Eliott Rosenberg and Trond Andersen and Rhine Samajdar and others},
year	= {2024},
URL	= {https://doi.org/10.1126/science.adi7877},
journal	= {Science},
pages	= {48-53},
volume	= {384},
  doi = {10.1126/science.adi7877},
  eprint = {2306.09333},
  archivePrefix = {arXiv}
}

@article{zhouEmergentStatisticalMechanics2019,
  title = {Emergent statistical mechanics of entanglement in random unitary circuits},
  author = {Zhou, Tianci and Nahum, Adam},
  year = {2019},
  journal = {Physical Review B},
  volume = {99},
  number = {17},
  pages = {174205},
  doi = {10.1103/PhysRevB.99.174205},
  eprint = {1804.09737},
  archivePrefix = {arXiv},
  url = {https://arxiv.org/abs/1804.09737},
  primaryClass = {cond-mat.stat-mech}
}

@article{baikAsymptoticsTracyWidomDistributions2008,
  title = {Asymptotics of {Tracy-Widom} distributions and the total integral of a {Painlev\'e II} function},
  author = {Baik, Jinho and Buckingham, Robert and DiFranco, Jeffery},
  year = {2008},
  journal = {Communications in Mathematical Physics},
  volume = {280},
  number = {2},
  pages = {463--497},
  doi = {10.1007/s00220-008-0433-5},
  eprint = {0704.3636},
  archivePrefix = {arXiv},
  url = {https://arxiv.org/abs/0704.3636}
}

@article{gopalakrishnanDistinctUniversalityClasses2024,
  title = {Distinct universality classes of diffusive transport from full counting statistics},
  author = {Gopalakrishnan, Sarang and Morningstar, Alan and Vasseur, Romain and Khemani, Vedika},
  year = {2024},
  journal = {Physical Review B},
  volume = {109},
  pages = {024417},
  doi = {10.1103/PhysRevB.109.024417},
  eprint = {2203.09526},
  archivePrefix = {arXiv},
  primaryClass = {cond-mat.stat-mech},
  url = {https://doi.org/10.1103/PhysRevB.109.024417}
}

@article{krajnikUniversalAnomalousFluctuations2024,
  title = {Universal anomalous fluctuations in charged single-file systems},
  author = {Krajnik, {\v{Z}}iga and Schmidt, Johannes and Pasquier, Vincent and Prosen, Toma{\v{z}} and Ilievski, Enej},
  year = {2024},
  journal = {Physical Review Research},
  volume = {6},
  pages = {013260},
  doi = {10.1103/PhysRevResearch.6.013260},
  eprint = {2208.01463},
  archivePrefix = {arXiv},
  primaryClass = {cond-mat.stat-mech},
  url = {https://doi.org/10.1103/PhysRevResearch.6.013260}
}

@article{takeuchiPartialYetDefinite2025,
  title = {Partial yet definite emergence of the {Kardar-Parisi-Zhang} class in isotropic spin chains},
  author = {Takeuchi, Kazumasa A. and Takasan, Kazuaki and Busani, Ofer and Ferrari, Patrik L. and Vasseur, Romain and De Nardis, Jacopo},
  year = {2025},
  journal = {Physical Review Letters},
  volume = {134},
  pages = {097104},
  doi = {10.1103/PhysRevLett.134.097104},
  eprint = {2406.07150},
  archivePrefix = {arXiv},
  primaryClass = {cond-mat.stat-mech},
  url = {https://doi.org/10.1103/PhysRevLett.134.097104}
}

@article{fisherRandomQuantumCircuits2023,
  title = {Random Quantum Circuits},
  author = {Fisher, Matthew P. A. and Khemani, Vedika and Nahum, Adam and Vijay, Sagar},
  year = {2023},
  journal = {Annual Review of Condensed Matter Physics},
  volume = {14},
  pages = {335--379},
  doi = {10.1146/annurev-conmatphys-031720-030658},
  eprint = {2207.14280},
  archivePrefix = {arXiv},
  url = {https://doi.org/10.1146/annurev-conmatphys-031720-030658}
}

@article{robertsChaosComplexityDesign2017,
  title = {Chaos and complexity by design},
  author = {Roberts, Daniel A. and Yoshida, Beni},
  year = {2017},
  journal = {Journal of High Energy Physics},
  volume = {2017},
  number = {4},
  pages = {121},
  doi = {10.1007/JHEP04(2017)121},
  eprint = {1610.04903},
  archivePrefix = {arXiv},
  url = {https://doi.org/10.1007/JHEP04(2017)121}
}

@article{bertiniScramblingRandomUnitary2020,
  title = {Scrambling in random unitary circuits: Exact results},
  author = {Bertini, Bruno and Piroli, Lorenzo},
  year = {2020},
  journal = {Physical Review B},
  volume = {102},
  pages = {064305},
  doi = {10.1103/PhysRevB.102.064305},
  eprint = {2004.13697},
  archivePrefix = {arXiv},
  url = {https://doi.org/10.1103/PhysRevB.102.064305}
}

@article{xuAccessingScramblingMatrix2020,
  title = {Accessing scrambling using matrix product operators},
  author = {Xu, Shenglong and Swingle, Brian},
  year = {2020},
  journal = {Nature Physics},
  volume = {16},
  pages = {199--204},
  doi = {10.1038/s41567-019-0712-4},
  eprint = {1802.00801},
  archivePrefix = {arXiv},
  url = {https://doi.org/10.1038/s41567-019-0712-4}
}

@article{xuLocalityQuantumFluctuations2019,
  title = {Locality, Quantum Fluctuations, and Scrambling},
  author = {Xu, Shenglong and Swingle, Brian},
  year = {2019},
  journal = {Physical Review X},
  volume = {9},
  pages = {031048},
  doi = {10.1103/PhysRevX.9.031048},
  eprint = {1805.05376},
  archivePrefix = {arXiv},
  url = {https://doi.org/10.1103/PhysRevX.9.031048}
}

@misc{hunterjonesOperatorGrowthRandom2018,
  title = {Operator growth in random quantum circuits with symmetry},
  author = {Hunter-Jones, Nicholas},
  year = {2018},
  eprint = {1812.08219},
  archivePrefix = {arXiv},
  primaryClass = {quant-ph},
  url = {https://arxiv.org/abs/1812.08219}
}

@article{collinsIntegrationRespectHaar2006,
  title = {Integration with Respect to the {Haar} Measure on Unitary, Orthogonal and Symplectic Group},
  author = {Collins, Beno{\^i}t and {\'S}niady, Piotr},
  year = {2006},
  journal = {Communications in Mathematical Physics},
  volume = {264},
  number = {3},
  pages = {773--795},
  doi = {10.1007/s00220-006-1554-3},
  eprint = {math-ph/0402073},
  archivePrefix = {arXiv},
  url = {https://doi.org/10.1007/s00220-006-1554-3}
}

@misc{hunterjonesUnitaryDesignsStatistical2019,
  title = {Unitary designs from statistical mechanics in random quantum circuits},
  author = {Hunter-Jones, Nicholas},
  year = {2019},
  eprint = {1905.12053},
  archivePrefix = {arXiv},
  primaryClass = {quant-ph},
  url = {https://arxiv.org/abs/1905.12053}
}

@article{jonayPhysicalTheoryTwoStage2024,
  title = {Physical theory of two-stage thermalization},
  author = {Jonay, Cheryne and Zhou, Tianci},
  year = {2024},
  journal = {Physical Review B},
  volume = {110},
  pages = {L020306},
  doi = {10.1103/PhysRevB.110.L020306},
  eprint = {2310.04491},
  archivePrefix = {arXiv},
  url = {https://doi.org/10.1103/PhysRevB.110.L020306}
}

@article{khemaniOperatorSpreadingEmergence2018,
  title = {Operator Spreading and the Emergence of Dissipative Hydrodynamics under Unitary Evolution with Conservation Laws},
  author = {Khemani, Vedika and Vishwanath, Ashvin and Huse, David A.},
  year = {2018},
  journal = {Physical Review X},
  volume = {8},
  pages = {031057},
  doi = {10.1103/PhysRevX.8.031057},
  eprint = {1710.09835},
  archivePrefix = {arXiv},
  url = {https://doi.org/10.1103/PhysRevX.8.031057}
}

@article{rakovszkyDiffusiveHydrodynamicsCorrelators2018,
  title = {Diffusive Hydrodynamics of Out-of-Time-Ordered Correlators with Charge Conservation},
  author = {Rakovszky, Tibor and Pollmann, Frank and von Keyserlingk, C. W.},
  year = {2018},
  journal = {Physical Review X},
  volume = {8},
  pages = {031058},
  doi = {10.1103/PhysRevX.8.031058},
  eprint = {1710.09827},
  archivePrefix = {arXiv},
  url = {https://doi.org/10.1103/PhysRevX.8.031058}
}

@article{nahumQuantumEntanglementGrowth2017,
  title = {Quantum Entanglement Growth under Random Unitary Dynamics},
  author = {Nahum, Adam and Ruhman, Jonathan and Vijay, Sagar and Haah, Jeongwan},
  year = {2017},
  journal = {Physical Review X},
  volume = {7},
  pages = {031016},
  doi = {10.1103/PhysRevX.7.031016},
  eprint = {1608.06950},
  archivePrefix = {arXiv},
  url = {https://doi.org/10.1103/PhysRevX.7.031016}
}

@article{maldacenaBoundChaos2016,
  title = {A bound on chaos},
  author = {Maldacena, Juan and Shenker, Stephen H. and Stanford, Douglas},
  year = {2016},
  journal = {JHEP},
  volume = {2016},
  number = {08},
  pages = {106},
  doi = {10.1007/JHEP08(2016)106},
  url = {https://doi.org/10.1007/JHEP08(2016)106}
}

@article{hosurChaosQuantumChannels2016,
  title = {Chaos in quantum channels},
  author = {Hosur, Pavan and Qi, Xiao-Liang and Roberts, Daniel A. and Yoshida, Beni},
  year = {2016},
  journal = {JHEP},
  volume = {2016},
  number = {02},
  pages = {004},
  doi = {10.1007/JHEP02(2016)004},
  url = {https://doi.org/10.1007/JHEP02(2016)004}
}

@article{guTwoWayApproachOutofTimeOrder2022,
  title = {A two-way approach to out-of-time-order correlators},
  author = {Gu, Yingfei and Kitaev, Alexei and Zhang, Pengfei},
  year = {2022},
  journal = {JHEP},
  volume = {2022},
  number = {03},
  pages = {133},
  doi = {10.1007/JHEP03(2022)133},
  url = {https://doi.org/10.1007/JHEP03(2022)133}
}

@article{harrowSeparationOutOfTimeOrderedCorrelation2021,
  title = {Separation of Out-Of-Time-Ordered Correlation and Entanglement},
  author = {Harrow, Aram W. and Kong, Linghang and Liu, Zi-Wen and Mehraban, Saeed and Shor, Peter W.},
  year = {2021},
  journal = {PRX Quantum},
  volume = {2},
  number = {2},
  pages = {020339},
  doi = {10.1103/PRXQuantum.2.020339},
  url = {https://doi.org/10.1103/PRXQuantum.2.020339}
}

@article{gonzalezalonsoOutOfTimeOrderedCorrelatorQuasiprobabilitiesRobustly2019,
  title = {Out-of-Time-Ordered-Correlator Quasiprobabilities Robustly Witness Scrambling},
  author = {Gonz{\'{a}}lez Alonso, Jos{\'{e}} Ra{\'{u}}l and Yunger Halpern, Nicole and Dressel, Justin},
  year = {2019},
  journal = {Physical Review Letters},
  volume = {122},
  number = {4},
  pages = {040404},
  doi = {10.1103/PhysRevLett.122.040404},
  url = {https://doi.org/10.1103/PhysRevLett.122.040404}
}

@misc{chenOperatorScramblingQuantum2018,
  title = {Operator scrambling and quantum chaos},
  author = {Chen, Xiao and Zhou, Tianci},
  year = {2018},
  eprint = {1804.08655},
  archivePrefix = {arXiv},
  primaryClass = {cond-mat.str-el},
  url = {https://arxiv.org/abs/1804.08655}
}

@article{aleinerMicroscopicModelQuantum2016,
  title = {Microscopic model of quantum butterfly effect: Out-of-time-order correlators and traveling combustion waves},
  author = {Aleiner, Igor L. and Faoro, Lara and Ioffe, Lev B.},
  year = {2016},
  journal = {Annals of Physics},
  volume = {375},
  pages = {378--406},
  doi = {10.1016/j.aop.2016.09.006},
  url = {https://doi.org/10.1016/j.aop.2016.09.006}
}

@article{khemaniVelocityDependentLyapunovExponents2018,
  title = {Velocity-dependent {Lyapunov} exponents in many-body quantum, semiclassical, and classical chaos},
  author = {Khemani, Vedika and Huse, David A. and Nahum, Adam},
  year = {2018},
  journal = {Physical Review B},
  volume = {98},
  number = {14},
  pages = {144304},
  doi = {10.1103/PhysRevB.98.144304},
  url = {https://doi.org/10.1103/PhysRevB.98.144304}
}

@article{chanSolutionMinimalModel2018,
  title = {Solution of a Minimal Model for Many-Body Quantum Chaos},
  author = {Chan, Amos and De Luca, Andrea and Chalker, J. T.},
  year = {2018},
  journal = {Physical Review X},
  volume = {8},
  pages = {041019},
  doi = {10.1103/PhysRevX.8.041019},
  url = {https://doi.org/10.1103/PhysRevX.8.041019}
}

@article{garrattLocalPairingFeynman2021,
  title = {Local Pairing of {Feynman} Histories in Many-Body {Floquet} Models},
  author = {Garratt, S. J. and Chalker, J. T.},
  year = {2021},
  journal = {Physical Review X},
  volume = {11},
  number = {2},
  pages = {021051},
  doi = {10.1103/PhysRevX.11.021051},
  url = {https://doi.org/10.1103/PhysRevX.11.021051}
}

@article{hoExactEmergentQuantum2022,
  title = {Exact Emergent Quantum State Designs from Quantum Chaotic Dynamics},
  author = {Ho, Wen Wei and Choi, Soonwon},
  year = {2022},
  journal = {Physical Review Letters},
  volume = {128},
  number = {6},
  pages = {060601},
  doi = {10.1103/PhysRevLett.128.060601},
  url = {https://doi.org/10.1103/PhysRevLett.128.060601}
}

@article{baoTheoryPhaseTransition2020,
  title = {Theory of the phase transition in random unitary circuits with measurements},
  author = {Bao, Yimu and Choi, Soonwon and Altman, Ehud},
  year = {2020},
  journal = {Physical Review B},
  volume = {101},
  pages = {104301},
  doi = {10.1103/PhysRevB.101.104301},
  url = {https://doi.org/10.1103/PhysRevB.101.104301}
}

@article{jianMeasurementInducedCriticalityRandom2020,
  title = {Measurement-induced criticality in random quantum circuits},
  author = {Jian, Chao-Ming and You, Yi-Zhuang and Vasseur, Romain and Ludwig, Andreas W. W.},
  year = {2020},
  journal = {Physical Review B},
  volume = {101},
  pages = {104302},
  doi = {10.1103/PhysRevB.101.104302},
  url = {https://doi.org/10.1103/PhysRevB.101.104302}
}

@article{mezeiMembraneTheoryEntanglement2018,
  title = {Membrane theory of entanglement dynamics from holography},
  author = {Mezei, M{\'{a}}rk},
  year = {2018},
  journal = {Physical Review D},
  volume = {98},
  pages = {106025},
  doi = {10.1103/PhysRevD.98.106025},
  url = {https://doi.org/10.1103/PhysRevD.98.106025}
}

@article{liEntanglementDomainWalls2023,
  title = {Entanglement Domain Walls in Monitored Quantum Circuits and the Directed Polymer in a Random Environment},
  author = {Li, Yaodong and Vijay, Sagar and Fisher, Matthew P. A.},
  year = {2023},
  journal = {PRX Quantum},
  volume = {4},
  pages = {010331},
  doi = {10.1103/PRXQuantum.4.010331},
  url = {https://doi.org/10.1103/PRXQuantum.4.010331}
}

@article{gongCoarseGrainedEntanglementOperator2022,
  title = {Coarse-Grained Entanglement and Operator Growth in Anomalous Dynamics},
  author = {Gong, Zongping and Nahum, Adam and Piroli, Lorenzo},
  year = {2022},
  journal = {Physical Review Letters},
  volume = {128},
  pages = {080602},
  doi = {10.1103/PhysRevLett.128.080602},
  url = {https://doi.org/10.1103/PhysRevLett.128.080602}
}

@article{sommersZeroTemperatureEntanglementMembranes2024,
  title = {Zero-temperature entanglement membranes in quantum circuits},
  author = {Sommers, Grace M. and Gopalakrishnan, Sarang and Gullans, Michael J. and Huse, David A.},
  year = {2024},
  journal = {Physical Review B},
  volume = {110},
  pages = {064311},
  doi = {10.1103/PhysRevB.110.064311},
  url = {https://doi.org/10.1103/PhysRevB.110.064311}
}

@article{mccullochFullCountingStatistics2023,
  title = {Full Counting Statistics of Charge in Chaotic Many-Body Quantum Systems},
  author = {McCulloch, Ewan and De Nardis, Jacopo and Gopalakrishnan, Sarang and Vasseur, Romain},
  year = {2023},
  journal = {Physical Review Letters},
  volume = {131},
  pages = {210402},
  doi = {10.1103/PhysRevLett.131.210402},
  url = {https://doi.org/10.1103/PhysRevLett.131.210402}
}

@article{sierantEntanglementGrowthMinimal2023,
  title = {Entanglement Growth and Minimal Membranes in $(d+1)$ Random Unitary Circuits},
  author = {Sierant, Piotr and Schir{\`{o}}, Marco and Lewenstein, Maciej and Turkeshi, Xhek},
  year = {2023},
  journal = {Physical Review Letters},
  volume = {131},
  pages = {230403},
  doi = {10.1103/PhysRevLett.131.230403},
  url = {https://doi.org/10.1103/PhysRevLett.131.230403}
}

@article{haRougheningTransitionQuantum2025,
  title = {Roughening Transition in Quantum Circuits},
  author = {Ha, Hyunsoo and Huse, David A. and Sommers, Grace M.},
  year = {2025},
  journal = {Physical Review Letters},
  volume = {135},
  pages = {160603},
  doi = {10.1103/gkwd-8477},
  url = {https://doi.org/10.1103/gkwd-8477}
}

@article{rakovszkySubBallisticGrowthRenyi2019,
  title = {Sub-ballistic Growth of {R{\'{e}}nyi} Entropies due to Diffusion},
  author = {Rakovszky, Tibor and Pollmann, Frank and von Keyserlingk, C. W.},
  year = {2019},
  journal = {Physical Review Letters},
  volume = {122},
  pages = {250602},
  doi = {10.1103/PhysRevLett.122.250602},
  url = {https://doi.org/10.1103/PhysRevLett.122.250602}
}

@article{jonaySlowThermalizationSubdiffusion2024,
  title = {Slow thermalization and subdiffusion in {U(1)} conserving {Floquet} random circuits},
  author = {Jonay, Cheryne and Rodriguez-Nieva, Joaquin F. and Khemani, Vedika},
  year = {2024},
  journal = {Physical Review B},
  volume = {109},
  pages = {024311},
  doi = {10.1103/PhysRevB.109.024311},
  url = {https://doi.org/10.1103/PhysRevB.109.024311}
}

@article{chenQuantumButterflyEffect2020,
  title = {Quantum butterfly effect in polarized {Floquet} systems},
  author = {Chen, Xiao and Nandkishore, Rahul M. and Lucas, Andrew},
  year = {2020},
  journal = {Physical Review B},
  volume = {101},
  pages = {064307},
  doi = {10.1103/PhysRevB.101.064307},
  url = {https://doi.org/10.1103/PhysRevB.101.064307}
}

@article{marvianRestrictionsRealizableUnitary2022,
  title = {Restrictions on realizable unitary operations imposed by symmetry and locality},
  author = {Marvian, Iman},
  year = {2022},
  journal = {Nature Physics},
  volume = {18},
  pages = {283--289},
  doi = {10.1038/s41567-021-01464-0},
  url = {https://doi.org/10.1038/s41567-021-01464-0}
}

@article{emersonPseudoRandomUnitaryOperators2003,
  title = {Pseudo-Random Unitary Operators for Quantum Information Processing},
  author = {Emerson, Joseph and Weinstein, Yaakov S. and Saraceno, Marcos and Lloyd, Seth and Cory, David G.},
  year = {2003},
  journal = {Science},
  volume = {302},
  number = {5653},
  pages = {2098--2100},
  doi = {10.1126/science.1090790},
  url = {https://doi.org/10.1126/science.1090790}
}

@article{oliveiraGenericEntanglementCan2007,
  title = {Generic Entanglement Can Be Generated Efficiently},
  author = {Oliveira, Roberto and Dahlsten, Oscar C. O. and Plenio, Martin B.},
  year = {2007},
  journal = {Physical Review Letters},
  volume = {98},
  pages = {130502},
  doi = {10.1103/PhysRevLett.98.130502},
  eprint = {quant-ph/0605126},
  archivePrefix = {arXiv},
  url = {https://doi.org/10.1103/PhysRevLett.98.130502}
}

@article{znidaricExactConvergenceTimes2008,
  title = {Exact convergence times for generation of random bipartite entanglement},
  author = {{\v{Z}}nidari{\v{c}}, Marko},
  year = {2008},
  journal = {Physical Review A},
  volume = {78},
  pages = {032324},
  doi = {10.1103/PhysRevA.78.032324},
  eprint = {0809.0554},
  archivePrefix = {arXiv},
  primaryClass = {quant-ph},
  url = {https://doi.org/10.1103/PhysRevA.78.032324}
}

\end{document}